\documentclass[a4paper,11pt]{article}
\usepackage{amsmath,amssymb,color,epsfig,cite}
\usepackage{graphicx}
\usepackage{subfigure}
\usepackage{setspace}
\usepackage{physics}
\usepackage{amsfonts}
\usepackage{mathtools}
\usepackage{physics}
\pdfoutput=1 

\usepackage{jheppub} 

\usepackage[T1]{fontenc} 

\newcommand{\be}{\begin{equation}}
\newcommand{\ee}{\end{equation}}
\newcommand{\bea}{\setlength\arraycolsep{2pt} \begin{eqnarray}}
\newcommand{\eea}{\end{eqnarray}}
\newcommand{\nn}{\nonumber}
\DeclareMathOperator{\sgn}{sgn}
\let\a=\alpha \let\b=\beta \let\g=\gamma \let\d=\delta 
    \let\k=\kappa
 \let\m=\mu \let\n=\nu \let\x=\xi \let\p=\pi 
 \let\t=\tau  \let\f=\phi  
\let\w=\omega         

  \let\S=\Sigma  \let\F=\Phi \let\Y=\Psi
 \let\W=\Omega 

\def\ft#1#2{{\textstyle{\frac{\scriptstyle #1}{\scriptstyle #2} } }}
\def\fft#1#2{{\frac{#1}{#2}}}

\def\0{{\sst{(0)}}}
\def\1{{\sst{(1)}}}
\def\2{{\sst{(2)}}}
\def\3{{\sst{(3)}}}
\def\4{{\sst{(4)}}}
\def\5{{\sst{(5)}}}
\def\6{{\sst{(6)}}}
\def\7{{\sst{(7)}}}
\def\8{{\sst{(8)}}}
\def\sst#1{{\scriptscriptstyle #1}}

\def\im{{{\rm i\,}}}

\title{ Kappa vacua: Enhancing the Unruh temperature}

\author[a]{Arash Azizi}

\affiliation[a]{{\it The Institute for Quantum Science and Engineering,
Texas A\&M University, College Station, TX 77843, USA}}

\emailAdd{sazizi@tamu.edu}

\abstract{We elaborate more on $\kappa$-mode, a mode that was found by a combination of Rindler modes in the right and left Rindler wedges with opposite sign norms. We find a relation between different kappa vacua, which is a generalization of the thermofield double state. However, a $\kappa$-vacuum can be written in terms of the Rindler vacuum as the thermofield double state, with a modified Unruh temperature of $T_{\kappa}=\frac{\hbar a}{2\pi c k_B}\,\kappa$. Consequently, when a uniformly accelerated observer with an acceleration $a$ is immersed in a $\kappa$-vacuum, they perceive a thermal bath. However, the temperature experienced by the observer is a modified Unruh temperature denoted as $T_{\kappa}$. Remarkably, the Unruh temperature can be enhanced by an arbitrary factor of $\kappa$.}

\begin{document} 
\maketitle
\flushbottom

\section{Introduction}

Quantum field theory (QFT) in curved spacetime{\footnote{For a recent review on mathematical background see \cite{witten22QFT}.} gives us intriguing results such as Hawking radiation \cite{Hawking75} and the Unruh effect \cite{Unruh76}. These phenomena have not only far-reaching consequences, where any consistent theory of quantum gravity should reproduce them at appropriate limits, but also appear in diverse research fields such as theoretical condensed matter \cite{Iorio:2011yz,Bhardwaj:2022sfw}, quantum optics \cite{Scully2003,Scully2006}, and quantum information \cite{bruschi2010unruh,Bradler:2008ce},  in addition to their original research program, high energy theory. The latter is especially reflected in the recent developments on resolving the Hawking information paradox \cite{Hawking76} which have been conducted by two groups in  \cite{Penington19,Almheiri2019,
west_coast,east_coast}.

The quintessence of Hawking radiation and the Unruh effect is the fact that generally there is not a unique vacuum in QFT. Fulling \cite{Fulling73} was one of the first scholars who realized this point. The canonical way of describing QFT is as follows:
\begin{itemize}
\item 
Consider a field theory of desired spin in a given   manifold. The curved spacetime is treated classically (no back reaction from quantum fields on  the background). The solutions of the equation of motion, i.e., field modes, can be derived from the Lagrangian of the theory in the curved background.
\item
The positive norm modes, with respect to the appropriate inner product, can be computed. The negative norm modes, hence, shall be  the complex conjugate of the positive norm ones.
\item
The quantum field can be expanded in terms of the modes, where the annihilation and creation operators are associated with the positive and negative norm modes respectively.
\item
The set of  positive norm modes is not unique, and hence, there are different sets of annihilation and creation operators, which can be related to each other by Bogoliubov transformations. Moreover, an  annihilation operator of one set can be a combination of  annihilation and creation operators of the other set. 
\item
The vacuum in the theory is defined as a state which is annihilated by all annihilation operators. In general  vacua of two sets of modes are distinct,  unless the annihilation operators of one set can be written in terms of just  annihilation operators (not annihilation and creation) of another one.
\end{itemize}

Unruh \cite{Unruh76} introduced a thought-provoking field mode, subsequently named Unruh mode, to explore the connection between the conventional Minkowski plane wave and Rindler modes. Although the Unruh mode's vacuum state is that of the Minkowski, its form bears resemblance to the Rindler mode. Furthermore, Unruh studied the behavior of a particle detector undergoing a uniform acceleration in the Minkowski vacuum. The particle detector was initially conceptualized by Unruh as a box with discrete energy levels \cite{Unruh76}, and later simplified further by DeWitt \cite{Einstein100}, by considering the first two energy levels of the box. As a result, it came to be known as the Unruh-DeWitt detector. The detector, starting from the ground state, gets excited and emits a photon in an Unruh mode. A peculiar characteristic of the mode is its predominant presence in the wedge opposite to the location of the detector, as highlighted by Unruh and Wald \cite{UnruhWald84}. More recently, two Unruh-DeWitt detectors were envisaged in  \cite{Svidzinsky21prl,Svidzinsky21prr} and it was shown despite  the apparent violation of causality, the latter is upheld.

In contrast to the more conventional QFT in Minkowski spacetime, where the plane wave is the commonly conceived mode, it is important to note that there exist an infinite number of distinct field modes. To properly associate the annihilation and creation operators with these modes, it becomes necessary to utilize the inner product. In a recent study \cite{Azizi2022kappashort}, we introduced a general mode called the $\kappa$-mode, which is formed by combining Rindler modes with opposite sign norms in the right and left Rindler wedges. One of the intriguing aspects of $\kappa$-modes is the existence of distinct vacua that are characterized by a real positive parameter $\kappa$. Notably, well-known vacua such as Minkowski and Rindler can be regarded as special cases of $\kappa$-vacuum for $\kappa=1$ and as $\k\rightarrow 0$, respectively. 

The thermofield double state \cite{Unruh76, Israel76} is ubiquitous in theoretical physics, e.g., it is a key ingredient of the AdS/CFT dictionary \cite{Maldacena97, Witten98,GKP}, as depicted in  Maldacena's proposal \cite{Maldacena2003eternal} about the duality of an eternal AdS black hole in the bulk and two copies of boundary CFT in the thermofield double state. Moreover, thermofield double state appears in quantum optics under the name of two-mode squeezed state \cite{scully_zubairy_1997}. 

By obtaining a relation between $\k$-vacuum and $\k'$-vacuum in (\ref{gentfd}), our central result, we generalize the usual thermofield double state. It is important to note that the pre-factor $\eta$, present in the exponential in (\ref{gentfd}), generally does not take the form of $e^{-\frac{\beta\Omega}{2}}$. As a consequence, it is not possible to express the $\kappa$-vacuum in terms of the $\kappa'$-vacuum using the conventional thermofield double state representation. A uniformly accelerated observer with an acceleration $a$ experiences the Minkowski vacuum as a thermal bath with the Unruh temperature given by\footnote{In this paper, we adopt the convention of using natural units where $\hbar = k_{B} = c = 1$.  However, when presenting the temperature, we reintroduce these constants to ensure consistency and clarity.} $T_U = \frac{\hbar a}{2\pi c k_B}$. Now, let's consider the scenario where the uniformly accelerated observer is immersed in a $\kappa$-vacuum. Surprisingly, the observer once again perceives a thermal bath, but this time with a modified temperature denoted as $T_{\kappa} = \frac{\hbar a}{2\pi c k_B} \, \kappa$.

The rest of the paper is organized as follows. In section \ref{sectioninnerprod} we study the Klein-Gordon inner product, and we show why the positive norm mode is associated with the annihilation operator by relating the inner product between modes to the commutation relations between operators. In section \ref{oppositenorm} we study the $\k$-mode and elaborate more on what was reported in \cite{Azizi2022kappashort}. In section \ref{samenorm} we show that it is not possible to get a mode by combining the same sign norm Rindler modes. In section \ref{sectiongentfd} we find a relation between kappa vacua. Specifically, by finding a relation between a kappa vacuum and the Rindler one, we show an accelerated observer in a $\k$-vacuum perceives a modified Unruh temperature of $T_{\k}=\fft{\hbar a}{2\p ck_B}\,\k$. Finally we conclude in the last section \ref{conclusion}.

\section{The Klein-Gordon inner product} \label{sectioninnerprod}

The Klein-Gordon inner product plays a fundamental role in distinguishing positive and negative norm modes, which are associated with annihilation and creation operators, respectively. One significant advantage of the inner product is that it remains constant in time.  In this section, we establish the interconnection between the inner product and the commutation relations. Additionally, we utilize the inner product to explicitly determine the Minkowski plane wave and Rindler modes. Moreover, we demonstrate a method for generating a new positive norm mode by combining a given positive norm mode with its complex conjugate.

\subsection{The definition and properties of the inner product} 

According to standard quantum field theory, a quantum field may be written as an infinite superposition of solutions of the equation of motions, i.e., modes, with operator coefficients which turn out to be annihilation and creation operators. The question then arises how to distinguish between modes associate to these operators. In other words, one may write a field as\footnote{For an explanation of the notions utilized in this paper, please refer to the appendix \ref{notation}.}
\be
\F(x)=\int d\W \left(\,\F(x, \W) \,a_{\W} +\F^*(x, \W) \,a^\dagger_{\W}\right)\,. \label{field}
\ee
To interpret $a_{\W}$ and $a^\dagger_{\W}$ as annihilation and creation operators respectively, they have to satisfy the standard commutation relation
\be
 [a_\W,a^\dagger_{\W'}]=\d(\W-\W')\,. \label{comrel-aadagger}
\ee

Therefore, it is crucial to choose a correct mode  $\F(x, \W)$ associated with the annihilation operator, while its complex conjugate $\F^*(x, \W)$ is associated with the creation one. The key concept for distinguishing these two modes is the inner product. For the Klein-Gordon scalar field the inner product reads
\begin{equation}
\left\langle \f_1, \f_2\right\rangle=
-\im \int_{\S}\, \sqrt{-g}\,
d\S^\m \left(\f_1^{*} \, \partial_\m \f_2
-\partial_\m  \f_1^{*}\, \f_2\right)\,, \label{innerprod}
\end{equation}
where $\S$ is the appropriate Cauchy hypersurface. Note, we have used the convention of $(-,+, \cdots, +)$ for the Minkowski metric. If one were to choose mostly negative signature for the metric, then there would be an overall minus sign difference, namely $\left\langle f, g\right\rangle=\im \int_{\S}\, \sqrt{-g}\,
d\S^\m \left(f^{*} \, \partial_\m g-\partial_\m  f^{*}\, g\right)$.

It is worthwhile here to note about different conventions on the Klein-Gordon inner product among the  early investigators. We consider four of them in the following:
\begin{itemize}

\item \textbf{Hawking} \cite{Hawking75}:
$
\left\langle\phi_{1}, \phi_{2}\right\rangle_{\text{H}}=\frac{\im}{2} \int_{\Sigma}\left[\phi_{1} \partial_{\mu} \phi_{2}^{*}-\left(\partial_{\mu} \phi_{1}\right) \phi_{2}^{*}\right] d S^{\mu}\,.
$

\item \textbf{DeWitt} \cite{DEWITT1975}:
$
\left\langle\phi_{1}, \phi_{2}\right\rangle_{\text{DeW}}=-\im
\int_{\Sigma}\left[\phi_{1}^{*} \partial_{\mu} \phi_{2}-\left(\partial_{\mu} \phi_{1}^{*}\right) \phi_{2}\right] d S^{\mu}\,.
$

\item \textbf{Wald} \cite{Wald:1975kc}: 
$
\left\langle\phi_{1}, \phi_{2}\right\rangle_{\text{Wald}}=\im \int_{\Sigma}\left[\phi_{1}^{*} \partial_{\mu} \phi_{2}-\left(\partial_{\mu} \phi_{1}^{*}\right) \phi_{2}\right] d S^{\mu}\,.
$

\item 
\textbf{Unruh and Wald} \cite{UnruhWald84}:
$
\left\langle\phi_{1}, \phi_{2}\right\rangle_{\text{UW}}=\frac{\im}{2} \int_{\Sigma}\left[\phi_{1}^{*} \partial_{\mu} \phi_{2}-\left(\partial_{\mu} \phi_{1}^{*}\right) \phi_{2}\right] d S^{\mu}\,.
$

\end{itemize}

The inner product is anti-linear in the second argument only in  Hawking's notation and is linear in the rest. While all of the above authors used the mostly positive signature for the metric, the sign of the inner product is  not consistent among them. Of course, there is a factor of one half discrepancy too.

Some useful relations in Klein-Gordon inner product are as follows:
\be
\left\langle f,\a g+\b h\right\rangle=\a\left\langle f,g\right\rangle
+\b \left\langle f,h\right\rangle\,, \quad \quad \quad
\left\langle f, g\right\rangle^*=\left\langle g, f\right\rangle\,,  \quad \quad \quad
\left\langle f^*, g^*\right\rangle=-\left\langle f, g\right\rangle^*\,. \label{ipproperty}
\ee

Note, the above Klein-Gordon ``inner product'' is not actually an inner product, strictly speaking, since 
\be
\left\langle f^*,f^*\right\rangle=-\left\langle f,f\right\rangle\,,
\ee
and therefore the positivity of inner product has been violated. However, this property is very crucial in distinguishing between positive  and negative norm modes. One may associate the annihilation operator to a positive norm mode, while the mode's complex conjugate, with a negative norm, is associated to the creation one. Actually, the inner product is defined so as to satisfy the following properties:
\be
\left\langle \F(x, \W),\F(x, \W')\right\rangle=[a_\W,a^\dagger_{\W'}]=\d(\W-\W')\,, \quad \quad
\left\langle \F(u, \W),\F^*(u, \W')\right\rangle=-\left[a_\Omega, a_ {\Omega^{\prime}}\right]=0\,. \label{innerprodrelations}
\ee
We prove the above relations in the next subsection. Also,  using (\ref{ipproperty}), then  (\ref{innerprodrelations})  implies 
\be
\left\langle \F^*(x, \W),\F^*(x, \W')\right\rangle=-\d(\W-\W')\,.
\ee

In a $D+1$ dimensional Minkowski spacetime with the usual convention for the component, i.e., $0$ and $i$ represent time and space components, and for a constant time Cauchy surface, one has $n_0=1$, $n_i=0$, $n^0=-1$, and $n^i=0$. Here $n_\m$ represents the unit normal vector to the manifold. Then (\ref{innerprod}) indicates
\begin{equation}
\left\langle f, g\right\rangle=\im \int\, \,
d^D x  \left(f^{*} \, \partial_t g-\partial_t  f^{*}\, g\right)\,. \label{innerprodmink}
\end{equation}
Here we have $d\S^\m=\d^{\m0} \,n^0 d^D x =-d^D x$ for $\m=0$ and zero for the rest of indices. Note mostly minus sign convention for the metric yields $n^0=1$, and hence to keep (\ref{innerprodmink}), one should start off from  $\left\langle f, g\right\rangle=\im \int_{\S}\, \sqrt{-g}\,
d\S^\m \left(f^{*} \, \partial_\m g-\partial_\m  f^{*}\, g\right)$ as we have emphasized. 

In $1+1$ Minkowski spacetime, the metric is $ds^2=-dt^2+dx^2=-du\,dv$. Here, we set $c=1$, and consider the $(-,+)$ convention for the metric. The light-cone coordinates are $u=t-x\,,v=t+x$. Thus for a manifold of constant $u$, one has $n^v=-2\,, n^u=0$; while for a manifold of constant $v$, one has $n^u=-2\,, n^v=0$. Since $\sqrt{-g}=\ft12$, then the inner product (\ref{innerprod}) becomes 
\begin{equation}
\left\langle f, g\right\rangle=\im \int_{-\infty}^{\infty} d v\left(f^{*} \frac{\partial}{\partial v} g-\frac{\partial}{\partial v} f^{*} g\right) \,, \quad
\left\langle f, g\right\rangle=\im \int_{-\infty}^{\infty} d u\left(f^{*} \frac{\partial}{\partial u} g-\frac{\partial}{\partial u} f^{*} g\right)\,,\label{innerprod(u,v)}
\end{equation}
where  a constant $u$, and a constant $v$ manifolds were chosen in the above relation respectively.

\subsection{The inner product and commutation relations}

Here in this subsection we show the connection between the inner products and the commutation relations. Namely,
\be
[a_\W,a^\dagger_{\W'}]=\left\langle \F(x, \W),\F(x, \W')\right\rangle\,, \qquad \qquad
\left[a_\Omega, a_ {\Omega^{\prime}}\right]
=- \left\langle \F(u, \W),\F^*(u, \W')\right\rangle\,.\label{innerproduct-comrel}
\ee
In order to prove the above relations, one may start from the following:
\be
a_{\W}=\left\langle \F(x, \W),\F(x)\right\rangle\,,
\quad \quad
a^\dagger_{\W}=-\left\langle \F^*(x, \W),\F(x)\right\rangle\,, \label{ann.creat.def}
\ee
where they can be found from the field mode expansion (\ref{field}), and using the properties of the inner product (\ref{ipproperty}). Next, we present a very useful lemma as follows.

\textbf{Lemma:} For any modes $f(u,\W)$ and $g(u,\W)$ one has the following relation:

\be
\left[\langle f(u, \Omega), {\Phi}(u)\rangle,\left\langle g\left(u^{\prime}, \Omega^{\prime}\right), {\Phi}\left(u^{\prime}\right)\right\rangle\right]
=-\left\langle f(u, \W),g^*(u, \W')\right\rangle\,.
\label{lemma}
\ee
The proof is given in the appendix \ref{appendixlemma}.

Having used the lemma (\ref{lemma}), one can now prove the above mentioned (\ref{innerproduct-comrel})  interconnection between  the commutation relations  and the inner products. Namely,
\bea
\left[a_\Omega, a^{\dagger}_ {\Omega^{\prime}}\right]
&=&\left[\langle\Phi(u, \Omega), {\Phi}(u)\rangle,-\left\langle\Phi^*\left(u^{\prime}, \Omega^{\prime}\right), {\Phi}\left(u^{\prime}\right)\right\rangle\right]
= \left\langle \F(u, \W),\F(u, \W')\right\rangle\,, \nn\\
\left[a_\Omega, a_ {\Omega^{\prime}}\right]
&=&\left[\langle\Phi(u, \Omega), {\Phi}(u)\rangle,\left\langle\Phi\left(u^{\prime}, \Omega^{\prime}\right), {\Phi}\left(u^{\prime}\right)\right\rangle\right]
=- \left\langle \F(u, \W),\F^*(u, \W')\right\rangle\,,
\eea
where we have used (\ref{ann.creat.def}) and the lemma (\ref{lemma}).

\subsection{Some examples: Minkowski plane wave and Rindler}
In this subsection, we derive Minkowski plane wave and Rindler positive norm modes by utilizing the inner product. It is more convenient to work with the light-cone coordinate $(u,v)$.  We consider a massless Klein-Gordon field in $1+1$ dimensions. The field equation is  $\Box \F=0$, or in terms of light-cone coordinates  $\ft{\partial}{\partial u}\ft{\partial}{\partial v}\F=0$, and can be solved simply by $\F(u,v)=\F(u)+\Y(v)$, where $\F(u)$ and $\Y(v)$ are general functions indicating right- and left-moving waves respectively. Consider the change of coordinates
\be
u=-\ft1a e^{-a(\t-\x)}\,, \qquad \qquad 
v=\ft1a e^{a(\t+\x)}\,. \label{uv-Mink-Rind}
\ee
One may call $(\t,\x)$ Rindler coordinates \cite{Rindler66} and the metric shall be written as $ds^2= e^{2 a\, \x}\,(-d \t^2 +d \x^2)$.   For positive (negative) $a$, we have $u<0$ ($u>0$) and $v>0$ ($v<0$)  and the associated region in spacetime diagram is called Rindler right (left) wedge.

Also since 
$\F(u,v)=\F(u)+\Psi(v)$, without loss of generality, we consider the right moving wave, i.e.,  $\F(u)$.

\subsubsection*{Minkowski plane wave}

Right moving  Minkowski plane wave reads 
$\F(u,\W)=f(\W)\,e^{-\im u\W}$. Imposing the Klein-Gordon inner product (\ref{innerprod(u,v)}) yields  
\begin{align}
\left\langle \F(u,\W), \F(u,\W')\right\rangle
&=\im f^*(\W)f(\W') \int_{-\infty}^{\infty} d u \left( 
e^{\im u\W} \frac{\partial}{\partial u}\,  e^{-\im u\W'}
-\frac{\partial}{\partial u} e^{\im u\W} \,  e^{-\im u\W'}\right) \\
&=f^*(\W)f(\W')(\W+\W') \int_{-\infty}^{\infty} d u \,
e^{\im u(\W-\W')} 
=4\p\W \abs{f(\W)}^2 \d(\W-\W')\,. \nn
\end{align}
Hence $\left\langle \F(u,\W), \F(u,\W')\right\rangle=\d(\W-\W')$ yields the important fact that  $\W$ should be a positive real number and $f(\W)=\fft1{\sqrt{4\p\W}}$. It is a well-known convention that the frequency $\nu>0$ is used for this mode. Therefore, the positive norm mode in right moving  Minkowski plane wave reads
\be
\F(u,\nu)=\fft1{\sqrt{4\p\nu}}\,e^{-\im u\nu}\,,  \quad \text{$\nu>0$}\,.
\ee

It is easy to check the inner product of a mode and its conjugate vanishes.

\subsubsection*{Rindler}

Rindler mode can be found in Rindler right ($u<0, v>0$), or left ($u>0, v<0$) wedges.  Again, considering the right moving wave, we may write the mode as 
\be 
\F(u,\W)=\theta(u)\,f(\W)\,u^{\im \W}\,, \qquad 
\F(u,\W)=\theta(-u)\,g(\W)\,(-u)^{\im \W}\,,
\ee
for the left and right wedges respectively. Here $\W$ is a real number, not necessarily a positive one. The inner product for the left Rindler wedge reads  
\begin{align}
\hspace{-0.3cm}\left\langle \F(u,\W), \F(u,\W')\right\rangle
&=\im f^*(\W)f(\W') \int_{-\infty}^{\infty} \theta(u) d u \left( 
u^{-\im \W} \frac{\partial}{\partial u}\,  u^{\im \W'}
-\frac{\partial}{\partial u} u^{-\im \W} \,  
u^{\im \W'}\right) \nn\\
&=-f^*(\W)f(\W')(\W+\W') \int_{-\infty}^{\infty}\theta(u)
d u \,u^{-\im (\W-\W')-1} \nn\\
&=-4\p\W \abs{f(\W)}^2 \d(\W-\W')\,, 
\end{align}
therefore, the orthonormality condition implies $\W<0$ and $f(\W)=\fft1{\sqrt{-4\p\W}}$.

Similarly for the right wedge, one has 
\begin{align}
\hspace{-0.3cm}\left\langle \F(u,\W), \F(u,\W')\right\rangle
&=\im g^*(\W)g(\W') \int_{-\infty}^{\infty} \theta(-u) d u \left( 
(-u)^{-\im \W} \frac{\partial}{\partial u}\,  (-u)^{\im \W'}
-\frac{\partial}{\partial u} (-u)^{-\im \W} \,  
(-u)^{\im \W'}\right) \nn\\
&=g^*(\W)g(\W')(\W+\W') \int_{-\infty}^{\infty}\theta(-u)
d u \,(-u)^{-\im (\W-\W')-1} \nn\\
&=4\p\W \abs{f(\W)}^2 \d(\W-\W')\,, 
\end{align}
therefore, the orthonormality condition implies $\W>0$ and $f(\W)=\fft1{\sqrt{4\p\W}}$. Consequently the positive norm Rindler mode for right moving wave is 
\begin{align}
&\text{Left wedge:} & \qquad  \F(u,\W)&=\theta(u)\,\fft1{\sqrt{4\p\W}}\,u^{-\im \W}\,, \quad& \text{$\W>0$}\,, \nn\\
&\text{Right wedge:} & \qquad 
\F(u,\W)&=\theta(-u)\,\fft1{\sqrt{4\p\W}}\,(-u)^{\im \W}\,, \quad& \text{$\W>0$}\,. \label{Rindmodes}
\end{align}

Note $\left\langle f^*,f^*\right\rangle=-\left\langle f,f\right\rangle$ implies the following relations for all real $\W$:
\bea
\left\langle \theta(u) u^{\im\W}, \theta(u) u^{\im\W'}\right\rangle &=&-4\p \W\,\d(\W'-\W)\,, \nn\\
\left\langle \theta(-u) (-u)^{\im\W}, \theta(-u) (-u)^{\im\W'}\right\rangle 
&=&4\p \W\,\d(\W'-\W)\,,\nn\\
\left\langle \theta(u) u^{\im\W}, \theta(-u) (-u)^{\im\W'}\right\rangle 
&=&0\,,\label{u,u,innerprod}
\eea
where the last relation is obvious since $\theta(u)\theta(-u)=0$.

\subsection{Combining positive and negative norm modes} \label{neg.+pos.general}

An approach to obtain new modes is by taking a linear combination of positive and negative norm modes, ensuring that the resulting new mode satisfies the inner product relations. Note that in what follows in this subsection, the general frequency $\Omega$, may be either positive or negative, and thus, the limits of integrals will be omitted.

The new mode $\Y(u,\W)$ may be defined as 
\be
\Y(u, \Omega)=\alpha(\Omega) \Phi(u, \Omega)+\beta(\Omega) {\Phi}^*(u, \Omega)\,, \label{neg+pos}
\ee
where $\F(u,\W)$ is the initial mode, and  satisfies the following inner product relations
\be
\left\langle \F(u, \W),\F(u, \W')\right\rangle=\d(\W-\W')\,, 
\qquad \qquad 
\left\langle \F(u, \W),\F^*(u, \W')\right\rangle=0\,. \label{innerorods}
\ee
Here $\a(\W)$ and $\b(\W)$ are general complex coefficients. The inner product for two modes $\Y(u,\W)$ reads
\begin{align} 
\left\langle \Y(u, \W),\Y(u, \W')\right\rangle =&
\left\langle\alpha(\Omega) \Phi(u, \Omega)+\beta(\Omega) {\Phi}^*(u, \Omega)
,\alpha(\Omega') \Phi(u, \Omega')+\beta(\Omega') {\Phi}^*(u, \Omega')\right\rangle \nn\\
=& \alpha^*(\Omega) \alpha\left(\Omega^{\prime}\right)\left\langle{\Phi}(u, \Omega), {\Phi}\left(u, \Omega^{\prime}\right)\right\rangle+\alpha^*(\Omega) \beta\left(\Omega^{\prime}\right)\left\langle\Phi(u, \Omega), {\Phi}^*\left(u, \Omega^{\prime}\right)\right\rangle \nn\\
&+\beta^*(\Omega) \alpha\left(\Omega^{\prime}\right)\left\langle{\Phi}^*(u, \Omega),
\Phi(u, \Omega^{\prime})\right\rangle
+\beta^*\left(\Omega) \beta\left(\Omega^{\prime}\right)\left\langle\Phi^*(u, \Omega), \Phi^*\left(u, \Omega^{\prime}\right)\right.\right\rangle
 \nn\\
=& \Big(\abs{\alpha(\Omega)}^2-\abs{\b(\Omega)}^2\Big)  \delta\left(\Omega-\Omega^{\prime}\right)\,.
\end{align}
Requiring the answer to be $\delta\left(\Omega-\Omega^{\prime}\right)$ yields the  following constraint:
\be 
\abs{\alpha(\Omega)}^2-\abs{\b(\Omega)}^2=1\,.
\ee
Furthermore, the inner product of a mode $\Y(u,\W)$ and its conjugate should be zero. Namely,
\bea
\left\langle \Y(u, \W),\Y^*(u, \W')\right\rangle &=& 
\left\langle\alpha(\Omega) \Phi(u, \Omega)+\beta(\Omega) {\Phi}^*(u, \Omega)
,\alpha^*(\Omega') \Phi^*(u, \Omega')
+\beta^*(\Omega') {\Phi}(u, \Omega')\right\rangle \nn\\
&=&\Big(\alpha^*(\Omega) \b^*\left(\Omega\right)-
\b^*(\Omega) \a^*\left(\Omega\right)
\Big)  \delta\left(\Omega-\Omega^{\prime}\right)\,.
\eea
However, this is identically zero and it does not impose any constraint on $\a(\W)$ and $\b(\W)$. Hence, the only constraint for the coefficients is $\abs{\alpha(\Omega)}^2-\abs{\b(\Omega)}^2=1$.

Now, to find the Bogoliubov transformation between the operators of the old and new modes, one can express the field in terms of the old and new modes and their corresponding creation and annihilation operators. Namely,
\be
\F(u)=\int d\W \left(\,\F(u, \W) \,a_{\W} +\F^*(u, \W) \,a^\dagger_{\W}\right)
=\int d\W \left(\,\Y(u, \W) \,b_{\W} +\Y^*(u, \W) \,b^\dagger_{\W}\right)
\,,\label{2fields}
\ee
where here $a_{\W}$ and $b_{\W}$ represent the annihilation operators for the old (i.e., $\F(u, \W)$) and new (i.e., $\Y(u, \W)$) modes respectively. Finding out  the inner product of $\left\langle \F(u, \W),\F(u)\right\rangle$ yields
\bea
a_{\W}&=&\left\langle \F(u, \W),\F(u)\right\rangle
=\int d\W' \left(\left\langle \F(u, \W),\Y(u, \W')\right\rangle \,b_{\W'} +
\left\langle \F(u, \W),\Y^*(u, \W')\right\rangle  \,b^\dagger_{\W'}\right) \nn\\
&=& \a(\W) \,b_{\W} +\b^*(\W)\, b^\dagger_{\W}\,, \label{a=b+b-dagger}
\eea
where we have used (\ref{neg+pos}) and (\ref{innerorods}). Note that in the above Bogoliubov transformation, unlike the well-known Bogoliubov transformation between Rindler and plane wave Minkowski, there is no summation over frequencies.

\section{Kappa mode: combining Rindler modes with opposite sign norms } \label{oppositenorm}
In section (\ref{neg.+pos.general}), we discussed a technique for finding new modes by combining positive and negative norm modes. We applied this method in (\ref{neg+pos}) by requiring both  old and new modes to be functions of $u$. This method can be used for the Rindler mode in a specific wedge, however, to find a similar relation as the thermofield double state, two distinct modes belong to different regions, e.g., right and left wedges, are needed.  In other words, in the thermofield double state, the existence of creation/annihilation operators in \textit{both} right \textit{and} left Rindler wedges enables the expression of the Minkowski vacuum in terms of the Rindler vacuum. As a matter of fact, the thermofield double state is widely known in the quantum optics community as a two-mode squeezed state, with the two modes referring to the right and left Rindler wedges.

Nevertheless, as Unruh demonstrated in his original work \cite{Unruh76}, it is possible to extend the method presented in (\ref{neg.+pos.general}) by permitting a mixing of modes that depends on both $u$ and $-u$. Specifically, in the case of Rindler modes, one may consider a combination of Rindler modes in the opposite wedges. There are two simple ways to proceed: by combining the same-signed norm modes of opposite wedges or by combining the opposite-signed norm modes of opposite wedges. In this section, we will address the latter, and in the following section, we will study the former. 

\subsection{Constructing Kappa modes}

The ansatz for the opposite sign norm is
\begin{equation}
\F(u,\W)=\a(\Omega)\theta(-u)(-u)^{\im \Omega}
+\b(\Omega)\theta(u) u^{\im \Omega}\,. \label{ansatz2}
\end{equation}
Note for positive (negative) $\W$, the first (second) term has positive norm, while the  second (first) term has negative norm.

The inner product of two modes reads
\begin{align}
\langle \F(u,\W),& \F(u,\W')\rangle \nn\\
&=\left\langle \a(\Omega)\theta(-u)(-u)^{\im \Omega}
+\b(\Omega)\theta(u) u^{\im \Omega},
\a(\Omega')\theta(-u)(-u)^{\im \Omega'}
+\b(\Omega')\theta(u) u^{\im \Omega'} \right\rangle \nn\\
&=
\a^*(\Omega)\a(\Omega')\left\langle\theta(-u)(-u)^{\im \Omega}, 
\theta(-u)(-u)^{\im \Omega'}
\right\rangle 
+\b^*(\Omega)\b(\Omega')\left\langle\theta(u) u^{\im \Omega}, 
\theta(u) u^{\im \Omega'}
\right\rangle \nn\\
&= 4\p\W\, \Big( \abs{\a(\W)}^2-\abs{\b(\W)}^2 \Big) \,\d(\W-\W')
\,. \label{ip5}
\end{align}
Therefore, the positivity of the norm indicates 
\be  
4\p\W\, \Big( \abs{\a(\W)}^2-\abs{\b(\W)}^2 \Big)=1\,. \label{constraint2}
\ee
Note $\W$ now can be either a  positive or a negative real number.

Furthermore, the inner product of the mode and its conjugate becomes
\begin{align}
\langle \F(u&,\W), \F^*(u,\W')\rangle  \label{ip6}\\
&=\left\langle \a(\Omega)\theta(-u)(-u)^{\im \Omega}
+\b(\Omega)\theta(u) u^{\im \Omega},
\a^*(\Omega')\theta(-u)(-u)^{-\im \Omega'}
+\b^*(\Omega')\theta(u) u^{-\im \Omega'} \right\rangle \nn\\
&=
\a^*(\Omega)\a^*(\Omega')\left\langle\theta(-u)(-u)^{\im \Omega}, 
\theta(-u)(-u)^{-\im \Omega'}
\right\rangle 
+\b^*(\Omega)\b^*(\Omega')\left\langle\theta(u) u^{\im \Omega}, 
\theta(u) u^{-\im \Omega'}
\right\rangle \,. \nn
\end{align}
By using (\ref{u,u,innerprod}) one has
\be
\left\langle \theta(u) u^{\im\W}, \theta(u) u^{-\im\W'}\right\rangle
=-\left\langle \theta(-u) (-u)^{\im\W}, \theta(-u) (-u)^{-\im\W'}\right\rangle 
=-4\p \W\,\d(\W+\W')\,. \label{innerprod4}
\ee
Thus the inner product (\ref{ip6}) then becomes
\begin{align}
\langle \F(u,\W),& \F^*(u,\W')\rangle= 4\p\W \,\d(\W+\W')\, \Big( \a^*(\Omega)\a^*(\Omega')-\b^*(\Omega)\b^*(\Omega') \Big)\,.\label{ip7}
\end{align}
This should be zero; however, in contrast to the previous case of subsection (\ref{neg.+pos.general}), since $\W$ and $\W'$ can be any real numbers, the inner product is not automatically vanishing, but (\ref{ip7}) imposes a constraint on $\a(\W)$ and $\b(\W)$. Namely,
\be
\Big( \a^*(\Omega)\a^*(\Omega')-\b^*(\Omega)\b^*(\Omega') \Big)\,\d(\W+\W') =
\Big( \a^*(\Omega)\a^*(-\Omega)-\b^*(\Omega)\b^*(-\Omega) \Big)\,\d(\W+\W')=0\,,
\ee
or simply,
\be
\a(\Omega)\a(-\Omega)-\b(\Omega)\b(-\Omega)=0\,,\label{constraint3}
\ee
for all $\W$.

Consequently, the inner product imposes two constraints. Actually, there are infinite number of functions can satisfy the constraints. For instance, let us write the complex-valued coefficient as follows
\be
\a(\Omega)=r(\W)\,e^{\im \theta(\W)}\,,\qquad \qquad\b(\Omega)=s(\W)\,e^{\im \f(\W)}\,,\label{gen.constraint1}
\ee
where $r(\W)$ and $s(\W)$ are positive real moduli. Therefore (\ref{constraint2}) indicates 
\be  
r^2(\W)-s^2(\W)=\ft1{4\p\W}\,, \label{gen.constraint2}
\ee
and (\ref{constraint3}) yields
\be
r(\Omega)r(-\Omega)-s(\Omega)s(-\Omega)=0\,,\qquad \qquad 
\theta(\W)+\theta(-\W)=\f(\W)+\f(-\W)\,.\label{gen.constraint3}
\ee
Writing (\ref{gen.constraint2}) for $\W$ and $-\W$ and combining them with (\ref{gen.constraint3}) simply results 
\be
r(\Omega)=s(-\Omega)\,,\label{gen.constraint4}
\ee
for all real $\W$. Hence the relation for $r(\W)$ is as follows
\be
r^2(\W)-r^2(-\W)=\ft1{4\p\W}\,.\label{gen.constraint5}
\ee

Moreover, any odd function of $\W$ could satisfy the equation (\ref{gen.constraint3}) for  phases $\theta(\Omega)$ and $\f(\Omega)$.

One can find  modes with a restriction (\ref{gen.constraint5}) for the moduli and $\theta(\W)+\theta(-\W)=\f(\W)+\f(-\W)$ for phases. However, in this paper, we ignore the phases of $\a(\Omega)$ and $\b(\Omega)$, and thus set $\a(\Omega)=r(\Omega)$ and $\b(\Omega)=s(\Omega)=r(-\Omega)=\a(-\Omega)$. Furthermore, an additional simplification will be implemented by considering the  following ansatz
\be
\a(\W)=\frac{e^{\fft{\p \Omega}{2\k}}}{\sqrt{8 \pi \Omega\, \sinh{(\fft{\p \Omega}{\k} )}}} \,,
\qquad \qquad
\b(\W)=\frac{e^{-\fft{\p \Omega}{2\k}}}{\sqrt{8 \pi \Omega\, \sinh{(\fft{\p \Omega}{\k} )}}}\,,
\ee
where $\k$ is an arbitrary positive real number. Since we have ignored the phases of $\alpha(\Omega)$ and $\beta(\Omega)$, then  $\frac{e^{\fft{\pm \p \Omega}{2\k}}}{\sqrt{8 \pi \Omega\, \sinh{(\fft{\p \Omega}{\k} )}}}$ is a real-valued function, and moreover $\Omega$ can be either positive or negative, then it follows that $\kappa$ must be a positive real number.

Consequently, one may write the following final result for the new mode:
\be
\F(u,\W,\k)= \fft1{{\sqrt{8 \pi \Omega\, \sinh{(\fft{\p \Omega}{\k} )}}}}
\left\{\theta(-u)(-u)^{\im \Omega}\, 
e^{\fft{\p \Omega}{2\k}}
+\theta(u) u^{\im \Omega}\, e^{-\fft{\p \Omega}{2\k}} \right\} \,. \label{modeopp}
\ee
Since it is classified by a positive number $\k$, we call it $\k$-mode. The field can be written as 
\be
\F(u)=\int_{-\infty}^{\infty} d\W \Big(\F(u,\W,\k) {\cal A}_{\W,\k} + \F^*(u,\W,\k) {\cal A}^{\dagger}_{\W,\k} \Big)\,,
\ee
where we have used ${\cal A}_{\W,\k}$ to denote the annihilation operator for a $\k$-mode with the  frequency $\W$ (Note $\W$ is any real number). Explicitly, using (\ref{modeopp}) one has
\bea
\F(u) &=&\theta(-u) \int_{-\infty}^{\infty} d\W 
 \fft1{{\sqrt{8 \pi \Omega\, \sinh{(\fft{\p \Omega}{\k} )}}}} 
\left\{(-u)^{\im \Omega}\, 
e^{\fft{\p \Omega}{2\k}}\, {\cal A}_{\W,\k}
+(-u)^{-\im \Omega}\, 
e^{\fft{\p \Omega}{2\k}}\, 
{\cal A}^{\dagger}_{\W,\k}\right\} \nn\\
&& +\theta(u) \int_{-\infty}^{\infty} d\W 
 \fft1{{\sqrt{8 \pi \Omega\, \sinh{(\fft{\p \Omega}{\k} )}}}} 
\left\{ u^{\im \Omega}\, e^{-\fft{\p \Omega}{2\k}}\, 
{\cal A}_{\W,\k} 
+ u^{-\im \Omega}\, e^{-\fft{\p \Omega}{2\k}}\, 
{\cal A}^{\dagger}_{\W,\k}\right\}\,. \label{kappafield} 
\eea

\subsection{Rindler and Unruh as special cases}

In this subsection we find  Rindler and Unruh modes as special cases of the $\k$-mode.
\subsection*{Rindler}

Let $\k\rightarrow 0$ in (\ref{modeopp}). For positive $\W$, the first term of (\ref{modeopp}) survives, indicating the Rindler mode in the right wedge.  Thus the $\k$-mode in this special case reads
\be
\F(u, \W > 0,\k\rightarrow 0)= \fft1{\sqrt{4 \pi \W}}
\theta(-u)\, (-u)^{ \im \W} \,. \label{RindRight}
\ee

Similarly, considering $\k\rightarrow 0$ with negative  $\W$ in (\ref{modeopp}), the second term of (\ref{modeopp}) survives, indicating the Rindler mode in the left wedge. Namely,
\be
\F(u, \W < 0,\k\rightarrow 0)= \fft1{\sqrt{4 \pi \abs{\W}}}
\theta(u)\, u^{ -\im \abs{\W}}\,, \label{RindLeft}
\ee
since $\abs{\W}=-\W$ for $\W<0$.

\subsection*{Unruh}

It is simple to observe that the Unruh mode is a special case of  $\k$-mode for $\k=1$ in (\ref{modeopp}). Namely, 
\be
\F(u,\W,\k=1)= \fft1{\sqrt{8 \pi \Omega\, \sinh{(\p \Omega)}}}
\left\{\theta(-u)(-u)^{\im \Omega}\, e^{\frac{\pi \Omega}{2}}
+\theta(u) u^{\im \Omega}\, e^{-\frac{\pi \Omega}{2}}\right\} \,. \label{UM}
\ee

\subsection{Bogoliubov transformation between $\k$-modes}
The goal is to find the Bogoliubov transformation between distinct $\k$-modes. Using (\ref{kappafield}), one has
\bea
\F(u) &=&\theta(-u) \int_{-\infty}^{\infty} d\W 
 \fft1{{\sqrt{8 \pi \Omega\, \sinh{(\fft{\p \Omega}{\k} )}}}} 
\left\{(-u)^{\im \Omega}\, 
e^{\fft{\p \Omega}{2\k}}\, {\cal A}_{\W,\k}
+(-u)^{-\im \Omega}\, 
e^{\fft{\p \Omega}{2\k}}\, 
{\cal A}^{\dagger}_{\W,\k}\right\} \nn\\
&& +\theta(u) \int_{-\infty}^{\infty} d\W 
 \fft1{{\sqrt{8 \pi \Omega\, \sinh{(\fft{\p \Omega}{\k} )}}}} 
\left\{ u^{\im \Omega}\, e^{-\fft{\p \Omega}{2\k}}\, 
{\cal A}_{\W,\k} 
+ u^{-\im \Omega}\, e^{-\fft{\p \Omega}{2\k}}\, 
{\cal A}^{\dagger}_{\W,\k}\right\} \nn\\
&=&\theta(-u) \int_{-\infty}^{\infty} d\W 
 \fft1{{\sqrt{8 \pi \Omega\, \sinh{(\fft{\p \Omega}{\k'} )}}}} 
\left\{(-u)^{\im \Omega}\, 
e^{\fft{\p \Omega}{2\k'}}\, {\cal A}_{\W,\k'}
+(-u)^{-\im \Omega}\, 
e^{\fft{\p \Omega}{2\k'}}\, 
{\cal A}^{\dagger}_{\W,\k'}\right\} \nn\\
&& +\theta(u) \int_{-\infty}^{\infty} d\W 
 \fft1{{\sqrt{8 \pi \Omega\, \sinh{(\fft{\p \Omega}{\k'} )}}}} 
\left\{ u^{\im \Omega}\, e^{-\fft{\p \Omega}{2\k'}}\, 
{\cal A}_{\W,\k'} 
+ u^{-\im \Omega}\, e^{-\fft{\p \Omega}{2\k'}}\, 
{\cal A}^{\dagger}_{\W,\k'}\right\}. \label{bogol1}
\eea
Next, we compare the factors of $\theta(-u) (-u)^{\im \W}$ and  $\theta(u) u^{\im \W}$ in (\ref{bogol1}). Comparing $\theta(-u) (-u)^{\im \W}$ factor indicates
\bea
&&\fft1{{\sqrt{8 \pi \Omega\, \sinh{(\fft{\p \Omega}{\k} )}}}} 
\left\{e^{\fft{\p \Omega}{2\k}}\, {\cal A}_{\W,\k}
+e^{-\fft{\p \Omega}{2\k}}\,{\cal A}^{\dagger}_{-\W,\k}\right\}\nn\\
&&
=\fft1{{\sqrt{8 \pi \Omega\, \sinh{(\fft{\p \Omega}{\k'} )}}}} 
\left\{e^{\fft{\p \Omega}{2\k'}}\, 
{\cal A}_{\W,\k'}
+e^{-\fft{\p \Omega}{2\k'}}\,
{\cal A}^{\dagger}_{-\W,\k'}\right\}\,. \label{bogol2}
\eea
One has to notice that since $-\infty < \W < \infty$, then $\theta(-u) (-u)^{\im \W}$ appears both in the first and second term of the first line of (\ref{bogol1}). Also, comparing $\theta(u) u^{\im \W}$ factors in (\ref{bogol1})  yields 
\bea
&&\fft1{{\sqrt{8 \pi \Omega\, \sinh{(\fft{\p \Omega}{\k} )}}}} 
\left\{e^{-\fft{\p \Omega}{2\k}}\, {\cal A}_{\W,\k}
+e^{\fft{\p \Omega}{2\k}}\,{\cal A}^{\dagger}_{-\W,\k}\right\}\nn\\
&&
=\fft1{{\sqrt{8 \pi \Omega\, \sinh{(\fft{\p \Omega}{\k'} )}}}} 
\left\{e^{-\fft{\p \Omega}{2\k'}}\, 
{\cal A}_{\W,\k'}
+e^{\fft{\p \Omega}{2\k'}}\,
{\cal A}^{\dagger}_{-\W,\k'}\right\}\,. \label{bogol3}
\eea

It is useful to find a transformation of  the pair 
$\begin{pmatrix}
{\cal A}_{\W,\k}\\
{\cal A}^{\dagger}_{-\W,\k}
\end{pmatrix}$. From (\ref{bogol2}) and (\ref{bogol3}), one has
\be
\begin{pmatrix}
{\cal A}_{\W,\k'}\\
{\cal A}^{\dagger}_{-\W,\k'}
\end{pmatrix} = 
\fft{\sgn{(\W)}}{\sqrt{{\sinh{(\fft{\p \Omega}{\k} )}\sinh{(\fft{\p \Omega}{\k'} )}}}}
\begin{pmatrix}
\sinh{\Big(\ft{\pi \W}2\,(\ft1{\k}+\ft1{\k'} \big)}\Big)\qquad
&\sinh{\Big(\ft{\pi \W}2\,(\ft1{\k'}-\ft1{\k} \big)}\Big)\\
\sinh{\Big(\ft{\pi \W}2\,(\ft1{\k'}-\ft1{\k} \big)}\Big)\qquad
&\sinh{\Big(\ft{\pi \W}2\,(\ft1{\k}+\ft1{\k'} \big)}\Big)
\end{pmatrix}\,
\begin{pmatrix}
{\cal A}_{\W,\k}\\
{\cal A}^{\dagger}_{-\W,\k}
\end{pmatrix}
\,. \label{bogolfinal}
\ee

The above relation is very crucial. It clearly shows, since the off-diagonal elements of the above matrix are non-vanishing for $\k\neq \k'$, that the annihilation operator in a mode $\k'$ depends upon both annihilation and creation operators of a mode $\k$, meaning that these modes have different vacua.

One may check explicitly the transformation between Rindler and Unruh operators. Namely, with $ \k=0,\,\k'=1$, one has 
\bea
{A}_{\W} &=&\fft1{\sqrt{1- e^{-2\p \W}}}\,
\Big(b_{R\W}- e^{-\p\W}\,b^{\dagger}_{L\W}\Big)\,,\nn\\
{A}_{-\W}&=&\fft1{\sqrt{1- e^{-2\p \W}}}\,
\Big(b_{L\W}- e^{-\p\W}\,b^{\dagger}_{R\W}\Big)\,,
\label{bogolUM-R}
\eea
where $\W>0$. This is in agreement with eqs (2.18) and (2.20) of the Unruh-Wald paper \cite{UnruhWald84}. Note here we adopt the following convention for Unruh and Rindler operators 
\begin{align}
{\cal A}_{\W,1} &={ A}_{\W}\,,
&\W& \in \mathbb{R}\,,    \nn\\
{\cal A}_{\W,0}&= b_{R\W}\,,\qquad 
{\cal A}_{-\W,0}=b_{L\W}\,,  &\W& \in \mathbb{R^+}
\,. \label{UM-Rindoperatordef}
\end{align}
where $A_\W$, $b_{R\W}$, and $b_{L\W}$ are the annihilation operators for Unruh, Rindler right wedge, and Rindler left wedge respectively.

One can get Rindler in terms of Unruh by either finding the inverse of (\ref{bogolUM-R}), or by setting  $ \k=1,\,\k'=0$ in (\ref{bogolfinal}). It reads
\bea
b_{R\W} &=&\fft1{\sqrt{1- e^{-2\p \W}}}\,
\Big({A}_{\W}+ e^{-\p\W}\,{A}^{\dagger}_{-\W}\Big)\,, \nn\\
b_{L\W}&=&\fft1{\sqrt{1- e^{-2\p \W}}}\,
\Big({A}_{-\W}+ e^{-\p\W}\,{A}^{\dagger}_{\W}\Big)\,,
\label{bogolR-UM}
\eea
again, recovering eq (2.24) of Unruh-Wald.

\subsection{Bogoliubov transformation between a  $\k$-mode and the Minkowski plane wave } \label{subsection:Bogol-kappa-Mink}
The goal is to find the Bogoliubov transformation between a $\k$-mode  and the Minkowski plane wave . Using (\ref{kappafield}), one has
\bea
\F(u) &=&\theta(-u) \int_{-\infty}^{\infty} d\W 
 \fft1{{\sqrt{8 \pi \Omega\, \sinh{(\fft{\p \Omega}{\k} )}}}} 
\left\{(-u)^{\im \Omega}\, 
e^{\fft{\p \Omega}{2\k}}\, {\cal A}_{\W,\k}
+(-u)^{-\im \Omega}\, 
e^{\fft{\p \Omega}{2\k}}\, 
{\cal A}^{\dagger}_{\W,\k}\right\} \nn\\
&& +\theta(u) \int_{-\infty}^{\infty} d\W 
 \fft1{{\sqrt{8 \pi \Omega\, \sinh{(\fft{\p \Omega}{\k} )}}}} 
\left\{ u^{\im \Omega}\, e^{-\fft{\p \Omega}{2\k}}\, 
{\cal A}_{\W,\k} 
+ u^{-\im \Omega}\, e^{-\fft{\p \Omega}{2\k}}\, 
{\cal A}^{\dagger}_{\W,\k}\right\} \nn\\
&=&\int_{0}^{\infty}
 \fft{d\n}{\sqrt{4 \p \n}}\, 
 \big(a_{\n}\,e^{-\im \n \, u}
 + a^\dagger_{\n}\,e^{\im \n \, u} \big)\,. \label{bogolGU-planewave1}
\eea
Next, one may find $a_{\n}$ by calculating  $\int_{-\infty}^{\infty}
du\, e^{\im \n \, u}\F(u)$. It reads
\bea
a_{\n} &=& \sqrt{\ft{\n}{\p}}  \int_{-\infty}^{\infty} d\W 
\fft{\im}{\sqrt{8 \pi \Omega\, \sinh{(\fft{\p \Omega}{\k} )}}} \label{bogolGU-planewave2}\\
&&\Bigg\{ 
-e^{\fft{\p \Omega}{2\k}}\, e^{\frac{\pi \Omega}{2}}\, \n^{-(1+\im \Omega)}\,\Gamma(1+\im \Omega)\,{\cal A}_{\W,\k}  
+e^{\fft{-\p \Omega}{2\k}}\, e^{-\frac{\pi \Omega}{2}}\, \n^{-(1+\im \Omega)}\,\Gamma(1+\im \Omega)\,{\cal A}_{\W,\k}  \nn\\
&&-e^{\fft{\p \Omega}{2\k}}\, e^{-\frac{\pi \Omega}{2}}\, \n^{-(1-\im \Omega)}\,\Gamma(1-\im \Omega)\,{\cal A}^{\dagger}_{\W,\k}
+e^{\fft{-\p \Omega}{2\k}}\, e^{\frac{\pi \Omega}{2}}\, \n^{-(1-\im \Omega)}\,\Gamma(1-\im \Omega)\,{\cal A}^{\dagger}_{\W,\k} \Bigg\} \nn
 \,, 
\eea
where we have used the following useful integrals:
\bea
\int_{-\infty}^{+\infty} d u\, e^{\im \n u}\, (-u)^{\im \Omega} \theta(-u) &=&\int_{0}^{\infty} d u\, e^{-\im \n u}\, u^{\im \Omega}=-\im \n^{-(1+\im \Omega)} e^{\frac{\pi \Omega}{2}} \Gamma(1+\im \Omega) \,, \nn\\
\int_{-\infty}^{+\infty} d u\, e^{\im \n u}\, u^{\im \Omega} \theta(u)&=&\int_{0}^{\infty} d u\, e^{\im \n u} u^{\im \Omega}=\im \n^{-(1+\im \Omega)} e^{-\frac{\pi \Omega}{2}} \Gamma(1+\im \Omega) \,.
\eea
Therefore, the final answer for the Bogoliubov transformation between $\k$-mode and plane wave Minkowski reads
\begin{align}
a_{\n} = \sqrt{\ft{\n}{\p}}  \int_{-\infty}^{\infty} d\W 
\fft{\im}{\sqrt{2 \pi \Omega\, \sinh{(\fft{\p \Omega}{\k} )}}} 
\Bigg\{&-\n^{-(1+\im \Omega)}\,\Gamma(1+\im \Omega)
\sinh{\Big(\ft{\pi \W}2\,(\ft1{\k}+1  \big)}\Big)
\,{\cal A}_{\W,\k}  \nn\\
&- \n^{-(1-\im \Omega)}\,\Gamma(1-\im \Omega)
\sinh{\Big(\ft{\pi \W}2\,(\ft1{\k}- 1  \big)}\Big)
\,{\cal A}^{\dagger}_{\W,\k} \Bigg\} 
 \,. \label{bogolGU-planewave3}
\end{align}

It is clear from the above relation that for $\k=1$ , i.e., the Unruh mode, the pre-factor of the creation operator vanishes. It  indicates the vacuum is the Minkowski one, as we have expected from the Unruh mode.

\subsection{Commutation relations for different $\k$}

Let's find the  commutation relation between different $\k$. To do so, one has to find the inner product of the modes with different $\k$. Using (\ref{modeopp}) we find the inner product between two positive norm modes as follows:
\begin{align}
\big\langle \F(u,\W,\k)&,\F(u,\W',\k')\big\rangle =
 \fft1{{\sqrt{8 \pi \Omega\, \sinh{(\fft{\p \Omega}{\k} )}}}}
 \fft1{{\sqrt{8 \pi \Omega'\, \sinh{(\fft{\p \Omega'}{\k'} )}}}}  \\\label{innerproddiffkappa1}
&\left\langle \theta(-u)(-u)^{\im \Omega}\, 
e^{\frac{\pi \Omega}{2\k}}
+\theta(u) u^{\im \Omega}\,
e^{\frac{-\pi \Omega}{2\k}},
\theta(-u)(-u)^{\im \Omega'}\, 
e^{\frac{\pi \Omega'}{2\k'}}
+\theta(u) u^{\im \Omega'}\,
e^{\frac{-\pi \Omega'}{2\k'}}
\right\rangle \nn\\
&=\fft{\sgn{(\W)}}{\sqrt{{\sinh{(\fft{\p \Omega}{\k} )}\sinh{(\fft{\p \Omega}{\k'} )}}}}
\,\sinh{\Big(\ft{\pi \W}2\,(\ft1{\k}+\ft1{\k'} \big)}\Big)\,
\d(\W-\W')\,, \nn
\end{align}
where we have used (\ref{u,u,innerprod}). Therefore the commutation relation between annihilation and creation operators with different $\k$ and $\k'$ is
\be
\left[{\cal A}_{\W, \k},
{\cal A}^\dagger_{\W', \k'}\right]=
\big\langle \F(u,\W,\k),\F(u,\W',\k')\big\rangle =
\fft{\sgn{(\W)}\,
\sinh{\Big(\ft{\pi \W}2\,(\ft1{\k}+\ft1{\k'} \big)}\Big)}{\sqrt{{\sinh{(\fft{\p \Omega}{\k} )}\sinh{(\fft{\p \Omega}{\k'} )}}}}\,\,
\d(\W-\W') \,.\label{commreldiffkappa1}
\ee
Note, the case of $\k=\k'$ yields the standard relation 
$\left[{\cal A}_{\W, \k},
{\cal A}^\dagger_{\W', \k}\right]=\d(\W-\W')$.

Next, the inner  product of the positive and negative norm modes with different $\k$ and $\k'$ reads 
\begin{align}
\big\langle \F(u,\W,\k)&,\F^*(u,\W',\k')\big\rangle =
 \fft1{{\sqrt{8 \pi \Omega\, \sinh{(\fft{\p \Omega}{\k} )}}}}
 \fft1{{\sqrt{8 \pi \Omega'\, \sinh{(\fft{\p \Omega'}{\k'} )}}}}  \\\label{innerproddiffkappa1}
&\left\langle \theta(-u)(-u)^{\im \Omega}\, 
e^{\frac{\pi \Omega}{2\k}}
+\theta(u) u^{\im \Omega}\,
e^{\frac{-\pi \Omega}{2\k}},
\theta(-u)(-u)^{-\im \Omega'}\, 
e^{\frac{\pi \Omega'}{2\k'}}
+\theta(u) u^{-\im \Omega'}\,
e^{\frac{-\pi \Omega'}{2\k'}}
\right\rangle \nn\\
&=\fft{\sgn{(\W)}}{\sqrt{{\sinh{(\fft{\p \Omega}{\k} )}\sinh{(\fft{\p \Omega}{\k'} )}}}}
\,\sinh{\Big(\ft{\pi \W}2\,(\ft1{\k}-\ft1{\k'} \big)}\Big)\,
\d(\W+\W')\,, \nn
\end{align}
where again (\ref{u,u,innerprod}) has been used. The commutation relation between annihilation operators with different $\k$ and $\k'$ is thus
\be
\left[{\cal A}_{\W, \k},
{\cal A}_{\W', \k'}\right]=
-\big\langle \F(u,\W,\k),\F^*(u,\W',\k')\big\rangle =
\fft{-\sgn{(\W)}\,
\sinh{\Big(\ft{\pi \W}2\,(\ft1{\k}-\ft1{\k'} \big)}\Big)}{\sqrt{{\sinh{(\fft{\p \Omega}{\k} )}\sinh{(\fft{\p \Omega}{\k'} )}}}}\,\,
\d(\W+\W') \,.\label{commreldiffkappa2}
\ee
Again, the case of $\k=\k'$ yields  
$\left[{\cal A}_{\W, \k},
{\cal A}_{\W', \k}\right]=0$, as it was expected.

It is interesting to note the Bogoliubov relation (\ref{bogolfinal}) can be found using the above commutation relation results. Providing that
$\begin{pmatrix}
{\cal A}_{\W,\k'}\\
{\cal A}^{\dagger}_{-\W,\k'}
\end{pmatrix}$ 
can be written in terms of a matrix multiplying 
$\begin{pmatrix}
{\cal A}_{\W,\k}\\
{\cal A}^{\dagger}_{-\W,\k}
\end{pmatrix}$
, the matrix elements can be found using the relations (\ref{commreldiffkappa1}) and  (\ref{commreldiffkappa2}). One then consequently  obtains (\ref{bogolfinal}) by employing the standard commutation relations 
$\left[{\cal A}_{\W, \k},
{\cal A}^\dagger_{\W', \k}\right]=\d(\W-\W')$, and 
$\left[{\cal A}_{\W, \k},
{\cal A}_{\W', \k}\right]=0$.

\section{Combining Rindler modes with same sign norms}\label{samenorm}
The same sign norm ansatz can be written as follows:
\begin{equation}
\F(u,\W)=\a(\Omega)\theta(-u)(-u)^{\im \Omega}
+\b(\Omega)\theta(u) u^{-\im \Omega}\,. \label{ansatz1}
\end{equation}
Note for a positive $\W$ both terms above have positive norm, while for a negative $\W$, both of them have negative one.

The new mode should satisfy the following inner products:
\be
\left\langle \F(u,\W), \F(u,\W')\right\rangle= \d(\W-\W')\,, 
\qquad \qquad
\left\langle \F(u,\W), \F^*(u,\W')\right\rangle=0\,. \label{ip1}
\ee
Hence, one may check 
\begin{align}
\langle \F(u,\W),& \F(u,\W')\rangle \nn\\
&=\left\langle\a(\Omega)\theta(-u)(-u)^{\im \Omega}
+\b(\Omega)\theta(u) u^{-\im \Omega},
\a(\Omega')\theta(-u)(-u)^{\im \Omega'}
+\b(\Omega')\theta(u) u^{-\im \Omega'}\right\rangle \nn\\
&=
\a^*(\Omega)\a(\Omega')\left\langle\theta(-u)(-u)^{\im \Omega}, 
\theta(-u)(-u)^{\im \Omega'}
\right\rangle 
+\b^*(\Omega)\b(\Omega')\left\langle\theta(u) u^{-\im \Omega}, 
\theta(u) u^{-\im \Omega'}
\right\rangle \nn\\
&= 4\p\W\, \Big( \abs{\a(\W)}^2+\abs{\b(\W)}^2 \Big) \,\d(\W-\W')
\,, \label{ip2}
\end{align}
where  we have used (\ref{u,u,innerprod}). The value of the above inner product, providing $\F(u,\W)$ is a positive norm mode, should be $\d(\W-\W')$, and hence
\be
4\p\W\, \Big( \abs{\a(\W)}^2+\abs{\b(\W)}^2 \Big) =1\,. \label{constraint1}
\ee
This implies  $\W$ should be positive.

Also, the inner product of the mode and its conjugate reads
\begin{align}
\langle &\F(u,\W), \F^*(u,\W')\rangle     \label{ip3} \\
&=\left\langle\a(\Omega)\theta(-u)(-u)^{\im \Omega}
+\b(\Omega)\theta(u) u^{-\im \Omega},
\a^*(\Omega')\theta(-u)(-u)^{-\im \Omega'}
+\b^*(\Omega')\theta(u) u^{\im \Omega'}\right\rangle \nn\\
&=
\a^*(\Omega)\a^*(\Omega')\left\langle\theta(-u)(-u)^{\im \Omega}, 
\theta(-u)(-u)^{-\im \Omega'}
\right\rangle 
+\b^*(\Omega)\b^*(\Omega')\left\langle\theta(u) u^{-\im \Omega}, 
\theta(u) u^{\im \Omega'}
\right\rangle\,, \nn
\end{align}
where by appropriate change of sign of $\W$ and $\W'$ in (\ref{u,u,innerprod}) one has
\be
\left\langle \theta(u) u^{-\im\W}, \theta(u) u^{\im\W'}\right\rangle
=\left\langle \theta(-u) (-u)^{\im\W}, \theta(-u) (-u)^{-\im\W'}\right\rangle 
=4\p \W\,\d(\W+\W')\,. \label{innerprod3}
\ee
Thus, the inner product (\ref{ip3}) becomes
\begin{align}
\langle \F(u,\W), \F^*(u,\W')\rangle   
=\Big(\a^*(\Omega)\a^*(\Omega')
+\b^*(\Omega)\b^*(\Omega')\Big) 4\p \W\,\d(\W+\W')\,.
\label{ip4}
\end{align}
The above term is zero automatically, since $\W$ and $\W'$ are both positive. Therefore, the inner product of  positive and negative norm modes  asserts no restriction on coefficients $\a(\W)$ and $\b(\W)$. Thus, (\ref{constraint1}) is the only constraint on $\a(\W)$ and $\b(\W)$. One may solve this constraint as follows:
\be
\a(\W)=\frac{e^{\fft{\p \Omega}{2\k}+\fft{\im \g}2}}{\sqrt{8 \pi \Omega\, \cosh{(\fft{\p \Omega}{\k} )}}} \,,
\qquad \qquad
\b(\W)=\frac{e^{-\fft{\p \Omega}{2\k}-\fft{\im \g}2}}{\sqrt{8 \pi \Omega\, \cosh{(\fft{\p \Omega}{\k} )}}}\,. \label{ansatzsamenorm}
\ee

Therefore, one may write the following final result for the mode:
\be
\F(u,\W,\k,\g)= \frac{1}{\sqrt{8 \pi \Omega\, 
\cosh{(\fft{\p \Omega}{\k} )}}}\,
\left\{\theta(-u)(-u)^{\im \Omega}\, 
e^{\fft{\p \Omega}{2\k}+\fft{\im \g}2}
+\theta(u) u^{-\im \Omega}\, 
e^{-\fft{\p \Omega}{2\k}-\fft{\im \g}2}\right\} \,. \label{modesame}
\ee

The field can be written as 
\be
\F(u)=\int_{0}^{\infty} d\W \Big(\F(u,\W,\k,\g) {\cal A}_{\W,\k,\g} + \F^*(u,\W,\k,\g) {\cal A}^{\dagger}_{\W,\k,\g} \Big)\,,
\ee
where we have used ${\cal A}_{\W,\k,\g}$ to denote the annihilation operator for the $\k$-mode. It is more convenient to drop $(\k,\g)$. Explicitly, using (\ref{modesame}), one has
\bea
\F(u) &=&\theta(-u) \int_{0}^{\infty} d\W 
\frac{1}{\sqrt{8 \pi \Omega\, 
\cosh{(\fft{\p \Omega}{\k} )}}}\, 
\left\{(-u)^{\im \Omega}\, 
e^{\fft{\p \Omega}{2\k}+\fft{\im \g}2}\,{\cal A}_{\W,\k,\g}
+(-u)^{-\im \Omega}\, 
e^{\fft{\p \Omega}{2\k}-\fft{\im \g}2}\,{\cal A}^{\dagger}_{\W,\k,\g}\right\} \nn\\
&&+ \theta(u) \int_{0}^{\infty} d\W 
\frac{1}{\sqrt{8 \pi \Omega\, 
\cosh{(\fft{\p \Omega}{\k} )}}}\, 
\left\{ u^{-\im \Omega}\,e^{-\fft{\p \Omega}{2\k}-\fft{\im \g}2}\, {\cal A}_{\W,\k,\g} 
+ u^{\im \Omega}\, 
e^{-\fft{\p \Omega}{2\k}+\fft{\im \g}2}\, {\cal A}^{\dagger}_{\W,\k,\g}\right\}\,. \nn\\ \label{fieldsame}
\eea

Please note that a key difference between the previous section and this one is the sign of $\Omega$. In the previous section, $\Omega$ could be either positive or negative, whereas in this section, (\ref{constraint1}) requires that $\Omega$ be positive. This constraint is the reason why we have chosen the ansatz (\ref{ansatzsamenorm}).

\subsection{Bogoliubov transformation between  different modes}
The goal is to find the Bogoliubov transformation between $\k$-modes with different $\k$ and $\g$. using (\ref{fieldsame}), one has
\bea
\F(u) &=&\theta(-u) \int_{0}^{\infty} d\W 
\frac{1}{\sqrt{8 \pi \Omega\, 
\cosh{(\fft{\p \Omega}{\k} )}}}\, 
\left\{(-u)^{\im \Omega}\, 
e^{\fft{\p \Omega}{2\k}+\fft{\im \g}2}\,{\cal A}_{\W,\k,\g}
+(-u)^{-\im \Omega}\, 
e^{\fft{\p \Omega}{2\k}-\fft{\im \g}2}\,{\cal A}^{\dagger}_{\W,\k,\g}\right\} \nn\\
&&+ \theta(u) \int_{0}^{\infty} d\W 
\frac{1}{\sqrt{8 \pi \Omega\, 
\cosh{(\fft{\p \Omega}{\k} )}}}\, 
\left\{ u^{-\im \Omega}\,e^{-\fft{\p \Omega}{2\k}-\fft{\im \g}2}\, {\cal A}_{\W,\k,\g} 
+ u^{\im \Omega}\, 
e^{-\fft{\p \Omega}{2\k}+\fft{\im \g}2}\, {\cal A}^{\dagger}_{\W,\k,\g}\right\} \nn\\ 
&=&\theta(-u) \int_{0}^{\infty} d\W 
\frac{1}{\sqrt{8 \pi \Omega\, 
\cosh{(\fft{\p \Omega}{\k'} )}}}\, 
\left\{(-u)^{\im \Omega}\, 
e^{\fft{\p \Omega}{2\k'}+\fft{\im \g'}2}\,{\cal A}_{\W,\k',\g'}
+(-u)^{-\im \Omega}\, 
e^{\fft{\p \Omega}{2\k'}-\fft{\im \g'}2}\,{\cal A}^{\dagger}_{\W,\k',\g'}\right\} \nn\\
&&+ \theta(u) \int_{0}^{\infty} d\W 
\frac{1}{\sqrt{8 \pi \Omega\, 
\cosh{(\fft{\p \Omega}{\k'} )}}}\, 
\left\{ u^{-\im \Omega}\,e^{-\fft{\p \Omega}{2\k'}-\fft{\im \g'}2}\, {\cal A}_{\W,\k',\g'} 
+ u^{\im \Omega}\, 
e^{-\fft{\p \Omega}{2\k'}+\fft{\im \g'}2}\, {\cal A}^{\dagger}_{\W,\k',\g'}\right\}\,. \nn\\ \label{bogol1same}
\eea
Comparing $\theta(-u) (-u)^{\im \W}$ and $\theta(u) u^{\im \W}$  in (\ref{bogol1same}) indicates
\bea
\fft1{\sqrt{8 \pi \W\, \cosh{(\fft{\p \Omega}{\k} )}}} 
\,e^{\fft{\p \Omega}{2\k}+\fft{\im \g}2}
\,{\cal A}_{\W,\k,\g} &=&
\fft1{\sqrt{8 \pi \W\, \cosh{(\fft{\p \Omega}{\k'} )}}} 
\,e^{\fft{\p \Omega}{2\k'}+\fft{\im \g'}2}
\,{\cal A}_{\W,\k',\g'}\,, \nn\\
\fft1{\sqrt{8 \pi \W\, \cosh{(\fft{\p \Omega}{\k} )}}} 
\,e^{-\fft{\p \Omega}{2\k}-\fft{\im \g}2}
\,{\cal A}_{\W,\k,\g} &=&
\fft1{\sqrt{8 \pi \W\, \cosh{(\fft{\p \Omega}{\k'} )}}} 
\,e^{-\fft{\p \Omega}{2\k'}-\fft{\im \g'}2}
\,{\cal A}_{\W,\k',\g'}\,. \label{bogol2same}
\eea
The above expressions simply yield $\k=\k'$ and $\g=\g'$. Consequently if there exists a mode in the same sign norm, then $\k$ and $\g$ would be unique.
 
\subsection{Bogoliubov transformation between the same sign norm and the opposite sign norm modes}

So far we have found if there were any mode in the same sign norm scenario, it would be just a unique $(\k,\g)$ mode. Here in this subsection, we find the Bogoliubov transformation between the latter mode and the previous case of the opposite sign norm mode. Using (\ref{kappafield}) and (\ref{fieldsame}) one has
\bea
\F(u) &=&\theta(-u) \int_{0}^{\infty} d\W 
\frac{1}{\sqrt{8 \pi \Omega\, 
\cosh{(\fft{\p \Omega}{\k} )}}}\, 
\left\{(-u)^{\im \Omega}\, 
e^{\fft{\p \Omega}{2\k}+\fft{\im \g}2}\,{\cal A}_{\W,\k,\g}
+(-u)^{-\im \Omega}\, 
e^{\fft{\p \Omega}{2\k}-\fft{\im \g}2}\,{\cal A}^{\dagger}_{\W,\k,\g}\right\} \nn\\
&&+ \theta(u) \int_{0}^{\infty} d\W 
\frac{1}{\sqrt{8 \pi \Omega\, 
\cosh{(\fft{\p \Omega}{\k} )}}}\, 
\left\{ u^{-\im \Omega}\,e^{-\fft{\p \Omega}{2\k}-\fft{\im \g}2}\, {\cal A}_{\W,\k,\g} 
+ u^{\im \Omega}\, 
e^{-\fft{\p \Omega}{2\k}+\fft{\im \g}2}\, {\cal A}^{\dagger}_{\W,\k,\g}\right\} \nn\\
&=&\theta(-u) \int_{-\infty}^{\infty} d\W 
 \fft1{{\sqrt{8 \pi \Omega\, \sinh{(\fft{\p \Omega}{\k'} )}}}} 
\left\{(-u)^{\im \Omega}\, 
e^{\fft{\p \Omega}{2\k'}}\, {\cal A}_{\W,\k'}
+(-u)^{-\im \Omega}\, 
e^{\fft{\p \Omega}{2\k'}}\, 
{\cal A}^{\dagger}_{\W,\k'}\right\} \nn\\
&& +\theta(u) \int_{-\infty}^{\infty} d\W 
 \fft1{{\sqrt{8 \pi \Omega\, \sinh{(\fft{\p \Omega}{\k'} )}}}} 
\left\{ u^{\im \Omega}\, e^{-\fft{\p \Omega}{2\k'}}\, 
{\cal A}_{\W,\k'} 
+ u^{-\im \Omega}\, e^{-\fft{\p \Omega}{2\k'}}\, 
{\cal A}^{\dagger}_{\W,\k'}\right\}\,. 
 \label{bogoloppsame}
\eea
Now, comparing  $\theta(u)u^{\im \W}$ for $\W>0$ in the above relation yields   
\be
\fft{e^{-\fft{\p \Omega}{2\k}+\fft{\im \g}2}}
{\sqrt{8 \pi \W\, \cosh{(\fft{\p \Omega}{\k} )}}}
{\cal A}^{\dagger}_{\W,\k,\g}
=\fft1{\sqrt{8 \pi \W\, \sinh{(\fft{\p \Omega}{\k'} )}}} 
\Big( e^{-\frac{\pi \W}{2\k'}}\,  {\cal A}_{\W,\k'} 
+e^{\frac{\pi \W}{2\k'}}\,  {\cal A}^{\dagger}_{-\W,\k'} 
\Big)\,. \label{oppsame1}
\ee
Also comparing  $\theta(-u)(-u)^{\im \W}$ for $\W>0$ in (\ref{bogoloppsame}) shows   
\be
\fft{e^{\fft{\p \Omega}{2\k}+\fft{\im \g}2}}
{\sqrt{8 \pi \W\, \cosh{(\fft{\p \Omega}{\k} )}}}
{\cal A}_{\W,\k,\g}
=\fft1{\sqrt{8 \pi \W\, \sinh{(\fft{\p \Omega}{\k'} )}}} 
\Big( e^{\frac{ \pi \W}{2\k'}}\,  {\cal A}_{\W,\k'} 
+e^{\frac{- \pi \W}{2\k'}}\,  {\cal A}^{\dagger}_{-\W,\k'} 
\Big)\,.\label{oppsame2}
\ee
Now it is clear while the left-hand sides of (\ref{oppsame2}) and Hermitian conjugate of (\ref{oppsame1}) are proportional, the right-hand sides are not. Therefore, one concludes the same sign norm mode cannot exist even for a unique value of $\k$ and $\g$.

Note the ansatz we have considered in (\ref{ansatzsamenorm}) is basically a simplest possible solution  satisfying the  restriction (\ref{constraint1}). However, even if a more general ansatz had been considered, there would still be no possibility for this combination to exist. The reason is primarily due to the requirement of positive values for the general frequency $\W$ in the same sign norm scenario. Due to this fact, even in a more complicated ansatz, if a specific term such as $\theta(u)u^{\im \W}$ ($\W>0$) is compared in the Bogoliubov transformation (\ref{bogoloppsame}) between the opposite sign norm and the same sign norm scenarios, there would be just one operator presents in the latter case, while both annihilation and creation operators present in the former case, since it allows both positive and negative general frequencies.   Following the same argument presented after eq. (\ref{oppsame2}) clearly shows the impossibility of the existence of the same sign norm scenario.

\section{$\k$-vacuum as a generalized non-thermofield double state} \label{sectiongentfd}
In 1976, Unruh \cite{Unruh76} and Israel \cite{Israel76} independently introduced the thermofield double state. One of the most well-known examples of the thermofield double state is the representation of the Minkowski vacuum in terms of the Rindler one. This relationship can be established through the Bogoliubov transformation connecting the Rindler modes and the Minkowski plane wave ones. Namely,
\be
\left(b_{L \w}-e^{-\frac{\pi \w}{a}} b_{R \w}^{\dagger}\right) \ket{0_{M}} =0\,, \quad \quad
\left(b_{R \w}-e^{-\frac{\pi \w}{a}} b_{L \w}^{\dagger}\right) \ket{0_{M}} =0\,. \label{bogolrindmink}
\ee
where $b_{R \w}$ and $b_{L \w}$ denoting the Rindler annihilation operators for the right and left wedges with a Rindler frequency $\w$ respectively. Also $a$ represents the constant acceleration of a particle. The  Minkowski vacuum is denoted by $\ket{0_{M}}$.

Then the Minkowski vacuum can be written in terms of entangled Rindler right-left wedges state as follows:
\be
\ket{0_{M}}=\fft1{\sqrt Z}\,
\exp{\int_0^{\infty} d\w\, e^{-\frac{\b \w}{2}}\,b_{L \w}^{\dagger}b_{R \w}^{\dagger}}\, \ket{0_{L}}\otimes\ket{0_{R}}\,, \label{tfd}
\ee
where $Z$ is the partition function, $\beta = \frac{1}{T} = \frac{2\pi}{\kappa a}$ is an inverse Unruh temperature, and $\ket{0_{R}}$ and $\ket{0_{L}}$ are the Rindler vacuum  in the right and left wedges respectively.

An important point should be  emphasized here. Consider two Rindler observers with acceleration $a$ and $a'$. There is a unique Rindler vacuum for both of these observers.  As calculated in the appendix \ref{appendixRindler}, this can be observed by finding the Bogoliubov transformation between two observers. It is
\be
\tilde b^{a'}_{R\, \w'}=(\ft{a}{a'})^{\fft12+\fft{\im \w}{a}}\, \tilde b^a_{R\, \w} 
\,, \qquad \qquad
\tilde b^{a'}_{L\, \w'}= (\ft{a}{a'})^{\fft12-\fft{\im \w}{a}} \,\tilde b^a_{L\, \w} \,, \label{Rind-a-a'}
\ee
with $\fft{\w}a=\fft{\w'}{a'}$. Here the superscript $a$ was inserted in $\tilde b^a_{R\, \w} \equiv b_{R \w} $ and $\tilde b^a_{L\, \w}\equiv b_{L \w}$ to emphasize that they are the annihilation operators for a Rindler  frequency $\w$ and an acceleration $a$  in the right and left wedges respectively.  The above is a trivial Bogoliubov transformation, i.e., the annihilation operator of the Rindler mode with acceleration $a$ is just proportional to the annihilation operator of the Rindler mode with acceleration $a'$ (\textit{not} both annihilation and creation operators). This  clearly indicates the uniqueness of the Rindler vacuum. With this consideration, one may notice the form of the thermofield double state in (\ref{tfd}) does not really depend upon the choice of $a$. One may exploit (\ref{Rind-a-a'}) to check this fact directly. 

\subsection{Relating different $\k$-vacua}

Distinct kappa vacua can be related to each other through the similar expression as stated in  Minkowski-Rindler relation. To obtain a relation between kappa vacua, first note ${\cal A}_{\W,\k'}\ket{0_{\k'}}=0$. Then by  exploiting the Bogoliubov transformation (\ref{bogolfinal}), one may observe 
\be
\Big({\cal A}_{\W,\k}-\eta_{\k,\k',\W}\, {\cal A}^{\dagger}_{-\W,\k}\Big) 
\ket{0_{\k'}}=0\,, \label{gentfdbogol}
\ee
where
\be
\eta_{\k,\k',\W}=\fft{\sinh{\Big(\ft{\pi \W}2\,(\ft1{\k}-\ft1{\k'} \big)}\Big)}
{\sinh{\Big(\ft{\pi \W}2\,(\ft1{\k}+\ft1{\k'} \big)}\Big)}\,. \label{eta}
\ee

Following the well-known steps that lead to the Minkowski-Rindler relation, $\k$ and $\k'$ vacua can be related as follows:
\be
\ket{0_{\k'}}=\fft1{\sqrt{Z_{\k\k'}}}\,\,
\exp{\int_{0}^{\infty} d\W\,\eta_{\k,\k',\W}\,{\cal A}^{\dagger}_{\W, \k}\,{\cal A}^{\dagger}_{-\W, \k}}\, \ket{0_{\k}}\,, \label{gentfd}
\ee
where $Z_{\k\k'}$ is the normalization factor which depends upon $\k$ and $\k'$. In the above relation,  although $\W$ can be both positive and negative, the  integral is carried out  over just positive frequencies. Alternatively,  one may include all positive and negative frequencies, however, a factor of one half should be included in the integral, since $\eta_{\k,\k',\W}=\eta_{\k,\k',-\W}$.

The relationship between different kappa vacua, as given by equation (\ref{gentfd}), can be considered as a generalization of the thermofield double state. Since the factor $\eta_{\k,\k',\Omega}$ appearing in the exponential term of equation (\ref{gentfd}) cannot be expressed in the form of the usual factor found in the thermofield double state, namely $e^{-\frac{\beta \omega}{2}}$ as shown in equation (\ref{tfd}).

\subsection{$\k$-vacuum in terms of Rindler}

Although we mentioned above that (\ref{tfd}) cannot be  represented as the thermofield double state, a surprising result emerges when the right hand side of (\ref{tfd}) is considered as $\k \rightarrow 0$, e.g., Rindler vacuum. Following the convention of (\ref{UM-Rindoperatordef}), and using (\ref{eta}), one finds $\eta=e^{-\fft{\p \W}{\k} }$, and thus
\be
\ket{0_{\k}}=\fft1{\sqrt{ Z_{\k}}}\,
\exp{\int_0^{\infty} d\W\, e^{-\fft{\p \W}{\k} }\,b_{L \W}^{\dagger}b_{R \W}^{\dagger}}\, \ket{0_{L}}\otimes\ket{0_{R}}\,.\label{kappaRind0}
\ee
Now, using (\ref{Rindapp5}) in the appendix \ref{appendixRindler}, and setting $\W=\fft{\w}a$, one may find the expression
\be
\ket{0_{\k}}=\fft1{\sqrt{ Z_{\k}}}\,
\exp{\int_0^{\infty} d\w\, e^{-\fft{\p \w}{\k a}}\,b_{L \w}^{\dagger}b_{R \w}^{\dagger}}\, \ket{0_{L}}\otimes\ket{0_{R}}\,, \label{kappaRind}
\ee
where again $\tilde{b}^a_{R\, \w}$ is simplified to $b_{R \w}$; likewise for the left wedge operators. 

The above expression is actually a thermofield double state with an inverse  temperature of $\beta = \frac{1}{T} = \frac{2\pi}{\kappa a}$. It should be noted that the uniqueness of the Rindler vacuum is crucial here. For instance, one might be tempted to think that (\ref{kappaRind}) is simply the Minkowski vacuum by removing $\k$ by means of  rescaling the acceleration, i.e., $a \rightarrow \frac{a}{\kappa}$. However, this is not the case. This is because the condition $\ft{\w}{a}=\ft{\w'}{a'}$ requires that the rescaling of $a$ to $a' = \ft{a}{\kappa}$ must be accompanied by a corresponding rescaling of $\omega$ to $\omega' = \ft{\w}{\kappa}$. Hence $e^{-\fft{\p \w}{\k a}}$ in (\ref{kappaRind}) remains unchanged, i.e., $e^{-\fft{\p \w}{\k a}} \rightarrow e^{-\fft{\p \w'}{\k a'}}$. Therefore, one cannot remove $\k$ in (\ref{kappaRind}) by a simple rescaling of the acceleration.

For retrieving the familiar case of the Minkowski vacuum in terms of the Rindler one, we should consider a  $\k$-vacuum as the Minkowski vacuum, namely $\k=1$. The  latter appears trivially by setting $\k=1$ in (\ref{kappaRind}).

\subsection{$\k$-vacuum in terms of Minkowski}
The last special case is finding a general $\k$-vacuum in terms of the  Minkowski vacuum. This can be observed readily by setting $\k=1$ in (\ref{gentfd}) and using the convention in (\ref{UM-Rindoperatordef}). Namely,
\be
\ket{0_{\k}}=\fft1{\sqrt{Z_{\k}}}\,\,
\exp{\int_{0}^{\infty} d\W\,\eta_{\k,\W}\, A^{\dagger}_{\W}\,A^{\dagger}_{-\W}}\, \ket{0_{M}}\,, \label{kappaMink}
\ee
where
\be
\eta_{\k,\W}=-\fft{\sinh{\Big(\ft{\pi \W}2\,(\ft1{\k}-1 \big)}\Big)}
{\sinh{\Big(\ft{\pi \W}2\,(\ft1{\k}+1 \big)}\Big)}\,, \label{etamink}
\ee
is a special case of (\ref{eta}) for $\k=1$ and relabeling $\k'$ as $\k$. Here as we have emphasized in (\ref{UM-Rindoperatordef}), $A_{\W}$ is an annihilation operator for the Unruh mode.

\section{Conclusion} \label{conclusion}

In this paper, we emphasize the significance of the Klein-Gordon inner product in properly defining a mode associated with annihilation and creation operators. Essentially, (\ref{innerprodrelations}) establishes a relationship between the inner product of modes and the commutation relations. Therefore, in order to satisfy the standard commutation relations, the inner product between modes should adhere to the form stated in (\ref{innerprodrelations}). Two well-known modes, namely the plane wave Minkowski and Rindler modes, have been analyzed as examples.

Furthermore, the inner product can be utilized to derive new modes from a given mode. In particular, we have developed this procedure for Rindler-like modes. In other words, by combining two Rindler modes in the right and left wedges, a new set of modes can be obtained. In this paper, we consider two cases: the combination of modes with the same sign of norm in the right and left wedges, and the combination of modes with opposite sign of norm in the right and left wedges. Interestingly, while the latter scenario yields an infinite set of modes parameterized by a real positive parameter $\k$, the former case does not yield any valid mode. The new $\k$-mode, inspired by the work of Unruh \cite{Unruh76}, gives rise to the Unruh and Rindler modes for special cases of $\k=1$ and $\k\rightarrow 0$, respectively.

By exploiting the Bogoliubov transformation in (\ref{bogolfinal}), one can find a relation, eq (\ref{gentfd}) between two kappa vacua. Namely, a $\k$-vacuum can be written in terms of another, say $\k'$-vacuum. The relation differs from that of the thermofield double in two aspects. Firstly, it features a modified coefficient in the exponential term, and secondly, it employs the $\k$-mode creation operators instead of the Rindler ones.  Surprisingly, when a general $\k'$-vacuum is expressed  in terms of  $\k \rightarrow 0$, e.g., Rindler vacuum, then the thermofield double state is recovered. However, it is important to note that the inverse temperature is modified to $\beta = \frac{1}{T} = \frac{2\pi}{\kappa a}$.

Consequently, a uniformly accelerated observer with a constant acceleration $a$ in a $\k$-vacuum still experiences a thermal bath. However, the Unruh temperature is modified to $T_{\kappa} = \frac{\hbar a}{2\pi c k_B} \, \kappa$.

\section*{Acknowledgments}

I am grateful to Girish Agarwal, Jonathan Ben-Benjamin, David Lee, Yusef Maleki, Anatoly Svidzinsky, and especially Marlan Scully and Bill Unruh for  illuminating discussions. I would like to thank Matthias Blau for his communication which improves the manuscript. I would like to thank Reed Nessler for his careful reading of the manuscript and providing helpful comments.

This work was supported by the
Robert A. Welch Foundation (Grant No. A-1261) and the National Science Foundation (Grant No. PHY-2013771).
\appendix 
\section{Notations} \label{notation}
In this paper, we have used the symbol $\W$, which, depending on the context, can be either $\W \in \mathbb{R}$ or $\W \in \mathbb{R^+}$, to represent a mode's general frequency. In the context of Rindler modes, the symbol $\w>0$ is used to denote the frequency when a specific acceleration is mentioned; on the other hand, if no specific acceleration is referred to, we use the symbol $\W>0$ for the frequency. For the Minkowski plane wave mode, the symbol $\n>0$ is used to denote the frequency.

In section \ref{oppositenorm}, ${\cal A}_{\W, \k}$ denotes the annihilation operator for a $\k$-mode with frequency $\W$. In  section \ref{samenorm}, ${\cal A}_{\W,\k,\g}$ is used to denote the   annihilation operator for a $(\k,\g)$-mode with a frequency $\W$. As will be explained at the end of this section, it is clear that no such mode exists.
\section{Proof of the lemma (\ref{lemma})} \label{appendixlemma}

To prove the lemma (\ref{lemma}), one may start from the definition of the inner product as follows 
\begin{align}
\left[\langle f(u, \Omega), {\Phi}(u)\rangle,\left\langle g\left(u^{\prime}, \Omega^{\prime}\right), {\Phi}\left(u^{\prime}\right)\right\rangle\right] \label{innerprod-comrel1}\\
&\hspace{-2.5cm}=-\int_{-\infty}^{+\infty} d u
\int_{-\infty}^{+\infty} d u' \Bigg\{
f^*(u, \Omega) g^*(u', \Omega')\, 
[\partial_u \F(u),\partial_{u'}\F(u')] \nn\\
&\hspace{1.6cm}- f^*(u, \Omega) \partial_{u'} g^*(u', \Omega')\, 
[\partial_u \F(u),\F(u')] \nn\\
&\hspace{1.6cm} 
 - \partial_u f^*(u, \Omega)  g^*(u', \Omega') 
[\F(u),\partial_{u'}\F(u')]  \nn\\
&\hspace{1.6cm}
+ \partial_u f^*(u, \Omega) \partial_{u'} g^*(u', \Omega')
[\F(u),\F(u')] \Bigg\}\,. \nn
\end{align}

There are four different commutation relations involved above where they are expressed as follows 
\begin{align}  
[\F(u),\F(u')] &= \ft{\im}4 \sgn(u'-u)\,, 
&[\partial_u \F(u),\F(u')] =&-\ft{\im}2 \d(u'-u)\,, \nn\\
[\F(u),\partial_{u'} \F(u')] &= \ft{\im}2 \d(u'-u)\,, 
&[\partial_u \F(u),\partial_{u'} \F(u')] =&\ft{\im}2 \d' (u'-u)\,. \label{comrel}
\end{align}
Next, one has to use the following properties of Dirac delta function
\bea
\int_{-\infty}^{+\infty} dx \, f(x) \delta(x) &=&f(0)\,, \label{diracdeltaprop}\\
\int_{-\infty}^{+\infty} dx \, f(x) \delta^{\prime}(x)&=&\int_{-\infty}^{+\infty} d x\left[\frac{d}{d x} \big(f(x) \delta(x)\big)
-\frac{d f}{d x} \delta(x)\right]= -\int_{-\infty}^{+\infty} d x f'(x) \d(x) \,.\nn 
\eea
Plugging (\ref{comrel}) and (\ref{diracdeltaprop}) in
 (\ref{innerprod-comrel1}), we have
\begin{align} 
\left[\langle f(u, \Omega), {\Phi}(u)\rangle,\left\langle g\left(u^{\prime}, \Omega^{\prime}\right), {\Phi}\left(u^{\prime}\right)\right\rangle\right] = \nn\\
&\hspace{-2cm}
-\frac{\im}{2} \int_{-\infty}^{+\infty} d u
\,f^*(u, \Omega) \partial_u g^*\left(u, \Omega^{\prime}\right) \nn\\
&\hspace{-2cm}
-\frac{\im}{2} \int_{-\infty}^{+\infty} d u 
\Big(f^*(u, \Omega) \partial_u 
g^* \left(u, \Omega^{\prime}\right)
-\partial_u f^*(u, \Omega) 
g^*\left(u, \Omega^{\prime}\right)\Big)  \nn\\
&\hspace{-2cm}
-\frac{\im}{4} \int_{-\infty}^{+\infty} d u du'\, \partial_u f^*(u, \Omega) \partial_{u'} g^*(u', \Omega') 
\, \sgn(u'-u)\,. \label{innerprod-comrel2}
\end{align}

To evaluate the last term above, one may proceed as follows
\begin{align}
\int_{-\infty}^{+\infty} d u du'\, \partial_u
f^*(u, \Omega) \partial_{u'} g^*(u', \Omega') 
\, \sgn(u'-u)  \nn\\
&\hspace{-7cm}
=\int_{-\infty}^{+\infty} du \Bigg\{-\int_{-\infty}^u 
d u^{\prime} \partial_u f^*\left(u, \Omega\right) 
\partial_{u'} g^*\left(u^{\prime}, \Omega^{\prime}\right) 
+\int_{u}^{\infty}d u^{\prime} \partial_u 
f^*\left(u, \Omega\right) \partial_{u'} 
g^*\left(u^{\prime}, \Omega^{\prime}\right) \Bigg\} \nn\\
&\hspace{-7cm}
=\int_{-\infty}^{+\infty} du \Bigg\{- \partial_u 
f^*\left(u, \Omega\right)
\Big(g^*\left(u, \Omega^{\prime}\right) 
-g^*\left(-\infty, \Omega^{\prime}\right) \Big) \nn\\
&\hspace{-7cm}
+\partial_u f^*\left(u, \Omega\right) 
\Big(  g^*\left(\infty, \Omega^{\prime}\right) 
-g^*\left(u, \Omega^{\prime}\right) \Big)\Bigg\} 
=-2\int_{-\infty}^{+\infty} du \, \partial_u
f^*\left(u, \Omega\right) g^*\left(u, \Omega^{\prime}\right) \,,
\end{align} 
where we have assumed the field is zero at infinities. Using the above relation, the commutation relation (\ref{innerprod-comrel2}) can be written finally as 
\begin{align} 
\left[\langle f(u, \Omega), {\Phi}(u)\rangle,\left\langle g\left(u^{\prime},\Omega^{\prime}\right), {\Phi}\left(u^{\prime}\right)\right\rangle\right]&   \nn\\
& \hspace{-3cm}=-\im \int_{-\infty}^{+\infty} d u
\Big(
f^*(u, \Omega) \partial_u g^*\left(u, \Omega^{\prime}\right) 
-\partial_u f^*(u, \Omega) g^*\left(u, \Omega^{\prime}\right) \Big) \nn\\
& \hspace{-3cm}=-\left\langle f(u, \W),g^*(u, \W')\right\rangle\,.
\label{innerprod-comrel3}
\end{align}
This completes the proof. 

\section{Rindler operators with different accelerations} \label{appendixRindler}
A relation between Rindler annihilation operators with different accelerations will be found in this appendix. To begin with, using the Rindler modes in (\ref{Rindmodes}), one may write down the field for the right moving wave as follows
\begin{align}
\F(u)= \int_0^{\infty} d\W\,\Bigg(
\theta(-u)\,\fft1{\sqrt{4\p\W}}\,(-u)^{\im \W}
b_{R \W}+
\theta(u)\,\fft1{\sqrt{4\p\W}}\,u^{-\im \W}
b_{L \W} + \text{h.c.} \Bigg)\,.
\end{align}
No acceleration appears in the above relation. However, one may use the  relations  in (\ref{uv-Mink-Rind}) to write the field as 
\begin{align}
\F(u)= \int_0^{\infty} \fft{d\W}{\sqrt{4\p\W}}\,\Bigg(
\theta(-u)\, a^{-\im \W}\,
e^{-\im \W a (\t-\x)}\,b_{R \W}
+\theta(u)\, a^{\im \W}\,
e^{-\im \W a (\t-\x)}\,
b_{L \W} + \text{h.c.} \Bigg)\,,
\end{align}
where $(\t,\x)$ is the Rindler coordinate. Here $a$ is a positive real number. Thus we have $u=-\ft1a e^{-a(\t-\x)}\,, v=\ft1a e^{a(\t+\x)}$ and $u=\ft1a e^{a(\t-\x)}\,,v=-\ft1a e^{-a(\t+\x)}$ for the right and left wedges respectively. Setting $\W=\ft{\w}a$ one may write
\begin{align}
\F_R(u) &= \int_0^{\infty} \fft{d\w}{\sqrt{4\p\w}}\,\Bigg( 
a^{-\ft12- \ft{\im\w}a}\,
e^{-\im \w (\t-\x)}\,b_{R \ft{\w}a} + \text{h.c.} \Bigg)\,, \nn\\
\F_L(u) &=\int_0^{\infty} \fft{d\w}{\sqrt{4\p\w}}
\,\Bigg( 
a^{-\ft12 +\ft{\im\w}a}\,
e^{-\im \w (\t-\x)}\,b_{L \ft{\w}a} + \text{h.c.} \Bigg)\,. \label{Rindapp3}
\end{align}

Alternatively, one may expand the field in terms of the usual plane wave mode, however, written in terms of the Rindler coordinate rather than the Minkowski one. Namely,  
\begin{align}
\F_R(u) = \int_0^{\infty} \fft{d\w}{\sqrt{4\p\w}}\,\big(
e^{-\im \w (\t-\x)}\,\tilde{b}^a_{R\, \w} + \text{h.c.} \big)\,, \quad
\F_L(u) =\int_0^{\infty} \fft{d\w}{\sqrt{4\p\w}}
\,\big( e^{-\im \w (\t-\x)}\,\tilde{b}^a_{L\, \w} + \text{h.c.} \big)\,. \label{Rindapp4}
\end{align}
The plane wave frequency is denoted as a Rindler frequency $\w$, and $\tilde{b}^a_{R\, \w}$ and $\tilde{b}^a_{L\, \w}$ are annihilation operators for the modes written in terms of specific acceleration $a$ and Rindler frequency $\w$. 

Therefore, comparing (\ref{Rindapp3}) and (\ref{Rindapp4}) above indicates 
\be
\tilde{b}^a_{R\, \w}=a^{-\ft12- \ft{\im\w}a}\,
b_{R\, \ft{\w}a} \,, \qquad \qquad
\tilde{b}^a_{L\, \w}=a^{-\ft12 +\ft{\im\w}a}\,
b_{L\, \ft{\w}a}\,. \label{Rindapp5}
\ee
We have to emphasize $b_{R\W}$ and $b_{L\W}$ are annihilation operators satisfying Rindler mode without specifying any accelerations. The modes and annihilation operators $b_{R\W}$ and $b_{L\W}$ in (\ref{Rindmodes}) were found via the Klein-Gordon inner product method. Since it has been derived without referring to a specific acceleration, one may call the modes in (\ref{Rindmodes}) as Rindler modes with general frequencies. On the other hand, $\tilde{b}^a_{R\, \w}$ and $\tilde{b}^a_{L\, \w}$ are annihilation operators for the modes written in terms of specific acceleration $a$. Both of these operators satisfy the canonical commutation relations. Namely, 
\be
[\tilde{b}^a_{R\, \w},
\tilde{b}^{a \dagger}_{R\, \w'}]=\d(\w-\w')\,,
 \qquad \qquad
[b_{R \W}, b^{\dagger}_{R\W'}]=\d(\W-\W')\,, \label{comrelRind}
\ee
and similar relations for the left wedge operators. Indeed (\ref{Rindapp5}) is consistent with the above commutation relations. 

With the above consideration, it is an easy task to compare the Rindler modes with different accelerations. Writing (\ref{Rindapp5}) with $(a',\w')$ indicates  
\be
\tilde{b}^{a'}_{R\, \w'}={a'}^{-\ft12- \ft{\im\w'}{a'}}\,
b_{R\, \ft{\w'}{a'}} \,, \qquad \qquad
\tilde{b}^{a'}_{L\, \w'}={a'}^{-\ft12 +\ft{\im\w'}{a'}}\,
b_{L\, \ft{\w'}{a'}}\,. \label{Rindapp5'}
\ee
Requiring the general frequency $\W=\ft{\w}a=\ft{\w'}{a'}$, one may observe
\be
b_{R\, \ft{\w}a}=b_{R\, \ft{\w'}{a'}}=a^{\ft12+ \ft{\im\w}a}\,\tilde{b}^a_{R\, \w}
=a'^{\ft12+ \ft{\im\w'}{a'}}\,\tilde{b}^{a'}_{R\, \w'}
 \,, \qquad 
b_{L\, \ft{\w}a}=b_{L\, \ft{\w'}{a'}}=
a^{\ft12 -\ft{\im\w}a}\,\tilde{b}^a_{L\, \w}
=a'^{\ft12 -\ft{\im\w'}{a'}}\,\tilde{b}^{a'}_{L\, \w'}
\,, \label{Rindapp6}
\ee
thus we have 
\be
\ft{\w}{a}=\ft{\w'}{a'}\,, \qquad \qquad
\tilde b^{a'}_{R\, \w'}=(\ft{a}{a'})^{\fft12+\fft{\im \w}{a}}\, \tilde b^a_{R\, \w} 
\,, \qquad \qquad
\tilde b^{a'}_{L\, \w'}= (\ft{a}{a'})^{\fft12-\fft{\im \w}{a}} \,\tilde b^a_{L\, \w} \,. \label{Rindapp7}
\ee
The above relation is a trivial Bogoliubov relation. Namely, annihilation operator for acceleration $a$ and Rindler frequency $w$ is proportional to that of the acceleration $a'$ and Rindler frequency $w'$. This fact indicates the Rindler vacuum is unique.

\bibliographystyle{jhep}
\bibliography{Kappa2Ver4}
\end{document}